\documentclass[reprint,amsmath,amssymb,superscriptaddress,aps,prl,floatfix]{revtex4-1}

\usepackage{graphicx}
\usepackage{dcolumn}
\usepackage{bm}
\usepackage[usenames,dvipsnames]{color}
\usepackage{epstopdf}
\usepackage{lipsum}
\newcommand{\Wen}{\mathit{We}}
\newcommand{\Ren}{\mathit{Re}}
\newcommand{\Ohn}{\mathit{Oh}}

\usepackage[colorlinks, citecolor=red]{hyperref}
\usepackage{siunitx}
\newcommand*\red{\textcolor{red}}

\begin{document}

\preprint{APS/123-QED}

\title{Impact forces of water drops falling on superhydrophobic surfaces}

\author{Bin Zhang}
\affiliation{%
	Department of Engineering Mechanics, AML, Tsinghua University, Beijing 100084, China
}%
\author{Vatsal Sanjay}
\affiliation{
	Physics of Fluids Group, Max Planck Center Twente for Complex Fluid Dynamics, MESA+ Institute, and J. M. Burgers Center for Fluid Dynamics, University of Twente, P.O. Box 217, 7500AE Enschede, Netherlands
}
\author{Songlin Shi}
\affiliation{%
	Department of Engineering Mechanics, AML, Tsinghua University, Beijing 100084, China
}%
\author{Yinggang Zhao}
\affiliation{%
	Department of Engineering Mechanics, AML, Tsinghua University, Beijing 100084, China
}%
\author{Cunjing Lv}
\email{cunjinglv@tsinghua.edu.cn}
\affiliation{%
	Department of Engineering Mechanics, AML, Tsinghua University, Beijing 100084, China
}%
\author{Xi-Qiao Feng}
\affiliation{%
	Department of Engineering Mechanics, AML, Tsinghua University, Beijing 100084, China
}%
\author{Detlef Lohse}
\affiliation{	
	Physics of Fluids Group, Max Planck Center Twente for Complex Fluid Dynamics, MESA+ Institute, and J. M. Burgers Center for Fluid Dynamics, University of Twente, P.O. Box 217, 7500AE Enschede, Netherlands
}
\affiliation{
	Max Planck Institute for Dynamics and Self-Organisation, Am Fassberg 17, 37077 G{\"o}ttingen, Germany
}
\date{\today}


\begin{abstract}
A falling liquid drop, after impact on a rigid substrate,  deforms and spreads, owing to the normal reaction force. Subsequently, if the substrate is non-wetting, the drop retracts and then jumps off. As we show here, not only is the impact itself associated with a distinct peak in the temporal evolution of the normal force, but also the jump-off, which was hitherto unknown. We characterize both peaks and elucidate how they relate to the different stages of the drop impact process. The time at which the second peak appears coincides with the formation of a Worthington jet, emerging through flow-focusing.
 Even low-velocity impacts can lead to a surprisingly high second peak in the normal force, even larger than the first one, namely when the  
 Worthington jet becomes singular due to the collapse of an air cavity in the drop.
\end{abstract}

\maketitle

In 1876, Arthur Mason Worthington \cite{worthington1877xxviii} published the first photographs of the drop impact process, stimulating artists and researchers alike for almost one-and-a-half centuries. Such drop impacts  on solid surfaces are highly relevant in inkjet printing \cite{lohse2022fundamental}, spray coating \cite{kim2007spray}, criminal forensics \cite{smith2018influence}, and many other industrial and natural processes \cite{Josserand2016,yarin2006drop,Yarin2017}. For most of these applications, the drop impact forces, which are the subject of this Letter, can lead to serious unwanted consequences, such as soil erosion \cite{Nearing1986} or the damage of engineered surfaces \cite{Ahmad2013, Amirzadeh2017, Gohardani2011}. A thorough understanding of the drop impact forces is thus needed to develop countermeasures against these damages \cite{cheng2021drop}. Consequently, recent studies analysed the temporal evolution of these forces \cite{Li2014, Soto2014, Philippi2016, Zhang2017, Gordillo2018, Mitchell2019, Zhang2019}. 
 
These studies were, however, up to now limited to wetting scenarios. Then, not surprisingly, the moment of the drop touch-down \cite{wagner1932stoss, Philippi2016} manifests itself in a pronounced peak in the temporal evolution of the drop impact force, whereas this force is much smaller during droplet spreading \cite{yarin2006drop, Wildeman2016}. For the non-wetting case, i.e. for superhydrophobic surfaces, the droplet dynamics is much richer: after reaching its maximal diameter, the drop recoils \cite{bergeron2001water} and can generate an upward, so-called  Worthington jet \cite{worthington1877xxviii, Bartolo2006Singular}. Ultimately, the drop can even rebound off the superhydrophobic surface \cite{Richard2002}. Such spectacular water repellency can occur  in nature \cite{onda1996super, Lafuma2003} and has technological applications \cite{Tuteja2007, Cho2016, Liu2017, Hao2016, Wu2020}, including on moving substrates \cite{zhan2021}, where the droplet dynamics is even richer. 
The feature of superhydrophobicity however is volatile and can fail due to external disturbance such as pressure \cite{Lafuma2003, Callies2005, Sbragaglia2007, Li2017}, evaporation \cite{Tsai2010, Chen2012, Papadopoulos2013},  mechanical vibration \cite{Bormashenko2007}, or the impact forces of prior droplets \cite{Bartolo2006Bouncing}.   

In this Letter, we extend the studies on drop impact forces to the impact on superhydrophobic surfaces. Our key result is that then, next to the first above-mentioned peak in the drop impact force at drop touch-down, a {\it second peak}  in the drop impact force occurs, which under certain conditions can be even more pronounced than the first peak. The physical origin of the second peak lies in momentum conservation: when at the final phase of droplet recoil the above-mentioned upward Worthington jet forms, momentum conservation also leads to a downward jet inside the drop \cite{lohse2004impact, Lohse2018, lee2020downward, mitra2021bouncing}. It manifests itself in the second peak in the temporal evolution of the force on the substrate. Using both experiments and direct numerical simulations (DNS) \cite{popinet-basilisk}, we will elucidate the physics of this very rich dynamical process and study its dependences on the control parameters.  

\begin{figure}
		\centering
		\includegraphics[width=\linewidth]{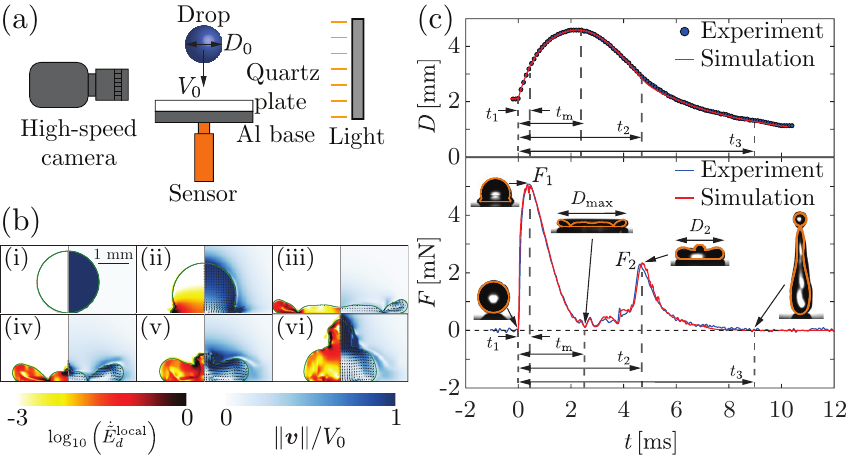}
		\caption{(a) Experimental setup: A water drop of diameter $D_0$ impacts on the superhydrophobic quartz plate at velocity $V_0$. (b) Numerical results for a drop impact dynamics for $D_0 = 2.05\,\si{\milli\meter}$ and $V_0 = 1.2\,\si{\meter}/\si{\second}$: $t =$ (i) $0\,\si{\milli\second}$ (touch-down), (ii) $0.37\,\si{\milli\second}$, (iii) $2.5\,\si{\milli\second}$, (iv) $3.93\,\si{\milli\second}$, (v) $4.63\,\si{\milli\second}$, and (vi) $5.25\,\si{\milli\second}$. The left part of each numerical snapshot shows the dimensionless local viscous dissipation rates $\dot{\tilde{E}}_d^{\text{local}}$ (see Supplemental Material, Eq.~\red{S8}) on a $\log_{10}$ scale and the right part the velocity field magnitude normalized with the impact velocity. The black velocity vectors are plotted in the center of mass reference frame of the drop to clearly elucidate the internal flow. (c)  Spreading diameter $D(t)$ and impact force $F(t)$ on the substrate as function of time: comparison between experiments  and simulations ($\Wen = 40$). The insets show representative snapshots at specific time instants overlaid with the drop boundaries from simulations in orange, revealing good agreement again. $F_1 \approx 5.1\,\si{\milli\newton}$ and  $F_2 \approx 2.3\,\si{\milli\newton}$ are the two peaks of the normal  force $F(t)$ at $t_1 \approx 0.37\,\si{\milli\second}$ and $t_2 \approx 4.63\,\si{\milli\second}$, respectively. $t_m$ is the moment corresponding to the maximum spreading of the drop and $t_3$ represents the end of contact ($F = 0$). See Supplemental Movie~\red{S1}.
		}
	\label{Fig:1}
\end{figure}

\noindent 
{\it Setup:}
The experimental setup is sketched in Fig.~\ref{Fig:1}(a). A water drop impacts a superhydrophobic substrate (see \cite{Li2017, Gauthier2015} for its preparation). We directly measure the  impact force $F(t)$ by synchronizing high-speed photo\-graphy with fast force sensing. In DNS, forces are calculated by integrating the pressure field at the substrate (see \cite{ramirez2020lifting} and Supplemental Material \S~\red{I} for details of the experimental and simulation setups \cite{supplMaterial}). The initial drop diameter $D_0$ ($2.05\,\si{\milli\meter} \le D_0 \le 2.76\,\si{\milli\meter}$) and the impact velocity $V_0$ ($0.38\,\si{\meter}/\si{\second} \le V_0 \le 2.96\,\si{\meter}/\si{\second}$) are independently controlled. The drop material properties are kept constant (density $\rho_d = 998\,\si{\kilogram}/\si{\meter}^{3}$, surface tension coefficient $\gamma = 73\,\si{\milli\newton}/\si{\meter}$, and dynamic viscosity $\mu_d = 1.0\,\si{\milli\pascal}\si{\second}$). All experiments were carried out at ambient air pressure and temperature. The Weber number (ratio of drop inertia to capillary pressure) $\Wen \equiv \rho_d V_0^2 D_0 / \gamma$ ranges between $1 - 400$ and the Reynolds number (ratio of inertial to viscous stresses) $\Ren \equiv \rho_d V_0 D_0 / \mu_d \approx 800\,\,\text{to}\,\,10^5$.  Note that for our simulations, we keep the drop Ohnesorge number (ratio of inertial-capillary to inertial-viscous timescales) $\Ohn \equiv \mu_d/\left(\rho_d\gamma D_0\right)^{1/2}$ constant at $0.0025$ to mimic $2\,\si{\milli\meter}$ diameter water drops. 

\noindent
{\it Formation of a second peak in the force and mechanism thereof:} 
Fig.~\ref{Fig:1}(b) illustrates the different stages of the drop impact process for $\Wen = 40$, and Fig.~\ref{Fig:1}(c) quantifies the spreading diameter $D(t)$ (the maximum width of the drop at time $t$) and the normal force $F(t)$ (see Supplemental Movie~\red{S1}). Note the remarkable quantitative agreement between the experimental and the numerical data for both $D(t)$ and $F(t)$, giving credibility to both. As the drop touches the surface [Fig.~\ref{Fig:1}(b-i)], the normal force $F(t)$ increases sharply to reach the first peak with amplitude $F_1 \approx 5.1\,\si{\milli\newton}$ in a very short time $t_1 \approx 0.37\,\si{\milli\second}$ [Fig.~\ref{Fig:1}(b-ii)]. At this instant, the spreading diameter $D(t)$ is equal to the initial drop diameter $D_0$, $D (t_1) \approx D_0$ \cite{Philippi2016, Zhang2017, Gordillo2018, Mitchell2019, Zhang2019}. Subsequently, the normal force reduces at a relatively slow rate to a minimum ($\approx 0\,\si{\milli\newton}$) at $t_m \approx 2.5\,\si{\milli\second}$. Meanwhile the drop reaches a maximum spreading diameter $D(t_m) = D_{max}$  [Fig.~\ref{Fig:1}(b-iii)]. The force profile $F(t)$, until this instant, is very close to that on a hydrophilic surface (see Supplemental Material \S~\red{II}). However, contrary to the wetting scenario, on superhydrophobic substrates, the drop starts to retract, creating high local viscous dissipation in the neck region connecting the drop with its rim [Fig.~\ref{Fig:1}(b-iii - b-iv)]. Through this phase of retraction, the normal reaction force is small, but shows several oscillations owing to traveling capillary waves for $2.5\,\si{\milli\second} < t < 3.8\,\si{\milli\second}$ [Fig.~\ref{Fig:1}(c)].  The drop retraction and the traveling capillary waves lead to flow focusing at the axis of symmetry, creating the Worthington jet [Fig.~\ref{Fig:1}(b-iv - b-v)] and hence also the opposite momentum jet that results in an increase in the normal force $F(t)$. Consequently, the hitherto unknown second peak appears, here with an amplitude $F_2 \approx 2.3\,\si{\milli\newton}$ and at time $t_2 \approx 4.63\,\si{\milli\second}$. Lastly, the normal force $F(t)$ decays slowly [Fig.~\ref{Fig:1}(b-v - b-vi)] to zero, here finally vanishing at $t_3 \approx 8.84\,\si{\milli\second}$. This time instant $t_3$ is a much better estimate for the drop contact time as compared to the one observed at complete detachment from side view images which is about $2\,\si{\milli\second}$ longer in this case \cite{Richard2002, chantelot_lohse_2021}. Therefore, in summary, here we have identified the mechanism for the formation of the second peak in the normal force and four different characteristic times,  $t_1$, $t_m$, $t_2$, and $t_3$ [Fig.~\ref{Fig:1} (c)]. 

\begin{figure}
	\centering
	\includegraphics[width=\linewidth]{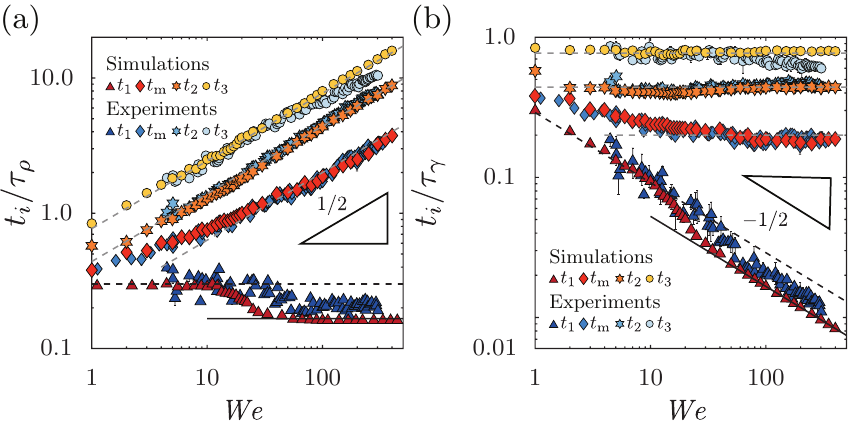}
 	\caption{Characteristic times as functions of $\Wen$. The times $t_1$, $t_m$, $t_2$, and $t_3$ are normalized by the inertial timescale $\tau_\rho = D_0/V_0$ (in (a)) or by the inertial-capillary timescale $\tau_\gamma = (\rho_d D_0^3/\gamma)^{1/2}$ (in (b)). The black dashed and solid lines represent $t_1 \approx 0.3\tau_\rho$ and $t_1 \approx 1/6\tau_\rho$, respectively. The gray dashed lines show  the best straight line fits to the experimental data, $t_m \approx 0.20\tau_\gamma$, $t_2 \approx 0.44\tau_\gamma$, and $t_3 \approx 0.78\tau_\gamma$.}
 	\label{Fig:2}
\end{figure}

\noindent 
{\it Weber-number dependence of the characteristics times:} 
Next, we look into the dependence of these times on the impact Weber number $\Wen$. The instant $t_1$ of the first peak of the force $F(t)$ scales with the inertial timescale [Fig.~\ref{Fig:2}(a)], i.e., $t_1 \sim \tau_\rho =  D_0/V_0$ with different $\Wen$-dependent prefactors ($\approx 0.3$ at low and $\approx 1/6$ at high $\Wen$, respectively). The solid black line in Fig.~\ref{Fig:2}(a) is the theoretical inertial limit, $t_1/\tau_\rho = 1/6$ \cite{Gordillo2018}, and matches our experimental and in particular numerical data. As seen from Fig.~\ref{Fig:2}, the other three characteristic times scale differently with $\Wen$ than $t_1$. Specifically, $t_2$ and $t_3$ become independent of $\Wen$ when rescaled with the inertial-capillary time $\tau_\gamma = \left(\rho_d D_0^3/\gamma\right)^{1/2}$ while $t_m$ has a weak $\Wen$-dependence at low $\Wen$, and becomes $\Wen$-independent only for $\Wen \gtrsim 10$, see Fig.~\ref{Fig:2}(b). The reason for this $\Wen$-independent behavior is that the impact process is analogous to one complete drop oscillation \cite{Richard2002} which is determined by the inertial-capillary time $\tau_\gamma$ \cite{rayleigh1879capillary}. Maximum spreading ($t_m$) occurs at almost one-quarter of a full oscillation (consistent with our result $t_m \approx 0.20 \tau_\gamma$) whereas the complete contact time $t_3$ takes about one full oscillation (consistent with our result $t_3 \approx 0.78 \tau_\gamma$). Finally, the time instant $t_2 \approx 0.44\tau_\gamma$ of the second peak in the impact force coincides with the time when the drop's motion changes from being predominantly radial to being vertical, as this moment is associated with the formation of the Worthington jet \cite[p. 18-20]{chantelot2018rebonds}. Note that here for the impact on  the superhydrophobic substrate, the duration of non-zero  forces (e.g.,  for $\Wen = 40$ we find $t_3/\tau_\rho \approx 5.2$  [Fig~\ref{Fig:1}(c)]) is much longer than that for the impact on a hydrophilic surfaces \cite{Gordillo2018, Mitchell2019}, where for the same $\Wen = 40$ one has $t_3/\tau_\rho \approx 2.0$.
    
\begin{figure}
	\centering
	\includegraphics[width=\linewidth]{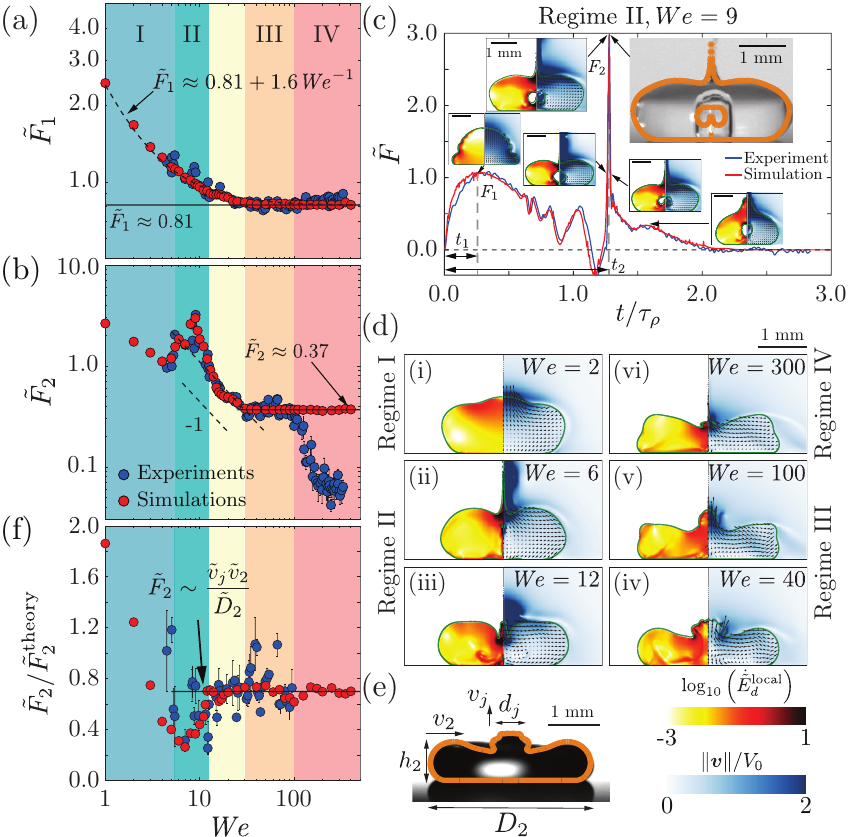}
	\caption{Dimensionless peak forces, (a) $\tilde{F}_1$, (b) $\tilde{F}_2$ as functions of $\Wen$. The variation of the second peak force $\tilde{F}_2$ with $\Wen$ divides the parameter space into four regimes: I. Capillary, II. Singular jet, III. Inertial, and IV. Splashing. (c) Evolution of the normal force $F(t)$ of an impacting drop for the case with highest $\tilde{F}_2$ ($\Wen = 9$). Note again the outstanding agreement between the experimental and the numerical results, including the various wiggles in the curve, which originate from capillary oscillations. Insets show representative snapshots at specific time instants. (d) Numerical snapshots at the instants of the second peak force ($t_2$) for $\Wen =$ (i) $6$, (ii) $12$, (iii) $40$, (iv) $100$, and (v) $300$. The left part of each numerical snapshot shows the dimensionless local viscous dissipation rates $\dot{\tilde{E}}_d^{\text{local}}$ (see Supplemental Material, Eq.~\red{S8}) on a $\log_{10}$ scale and the right part the velocity field magnitude normalized with the impact velocity. (e) Experimental drop geometry at $t_2$ for $\Wen = 40$ (along with the drop contour from numerics in orange) to illustrate its spreading diameter $D_2$, height $h_2$, retraction velocity $v_2$, jet diameter $d_j$ and jet velocity $v_j$. (f) Comparison of the second peak force $\tilde{F}_2$ with its theoretical prediction $\tilde{F}_2^{\text{theory}} = \tilde{v}_j\tilde{v}_2/\tilde D_2$ [Eq.~(\ref{Eqn::TheoryForce2})].
	}
	\label{Fig:3}
\end{figure}

\noindent 
{\it Weber-number dependence of the magnitude of the first peak:} 
As the drop falls on a substrate, momentum conservation implies $F_1 \sim V_0(\mathrm{d}m/\mathrm{d}t)$, where the mass flux $\mathrm{d}m/\mathrm{d}t$ can be calculated as $\mathrm{d}m/\mathrm{d}t \sim \rho_d V_0D_0^2$ \cite{Soto2014}. As a result, $F_1 \sim \rho_d V_0^2D_0^2$, as shown in Fig.~\ref{Fig:3}(a) for high Weber numbers ($\Wen > 30$, $F_1 \approx 0.81\rho_d V_0^2D_0^2$). This asymptote also matches the experimental and theoretical results of similar studies conducted on hydrophilic substrates \cite{Zhang2017, Gordillo2018}. Indeed, the first peak force originates from an inertial shock following the impact of drops onto an immobile substrate and is independent of the wettability. Further, the minimum Reynolds number for the current work is $800$, which is well above the criterion ($\Ren > 200$) for viscosity-independent results \cite{Zhang2017, Gordillo2018}. One would expect $\tilde{F}_1 \equiv F_1/\rho_d V_0^2D_0^2$ to be constant throughout the range of our parameter space. Nonetheless, when $\Wen < 30$, the data deviates from the inertial asymptote. Such deviations have been reported previously on hydrophilic surfaces as well \cite{Soto2014}. Here, inertia is not the sole governing force, and it competes with surface tension. We propose a generalization of the first peak of the impact force to $F_1 = \alpha_1\rho_d V_0^2 D_0^2 + \alpha_2(\gamma/ D_0) D_0^2$, based on dimensional analysis, with $\alpha_1$ and $\alpha_2$ as free parameters. From the best fit to all the experimental and numerical data, we obtain $\tilde{F}_1 \approx 0.81 + 1.6\Wen^{-1}$, which well describes the data, see Fig.~\ref{Fig:3}(a).

\vspace{1.2mm} 
\noindent
{\it Weber-number dependence of the magnitude of the second peak:} 
We now focus on the second peak $F_2$ of the impact force $F(t)$. In Fig.~\ref{Fig:3}(b), we show the $\Wen$-dependence of the non-dimensional version thereof, $\tilde{F}_2 \equiv F_2/(\rho_d V_0^2 D_0^2)$. We identify four main regimes, namely I. Capillary ($\Wen < 5.3$), II. Singular jet ($5.3 < \Wen < 12.6$), III. Inertial ($30 < \Wen < 100$), and IV. Splashing ($\Wen > 100$). The range $12.6 < \Wen < 30$ marks the transition from the singular jet to the inertial regime. In regime I ($\Wen < 4.5$), the amplitudes of both peaks, $F_1$ and $F_2$, are smaller than the resolution ($0.5\,\si{\milli\newton}$) of our force transducer, so we cannot extract the temporal variation of the normal reaction force from the experiments.
Capillary oscillations dominate the flow in this regime \cite{Richard2000}, leading to more than two peak forces, remarkably perfectly identical to what is observed in our simulations (see Fig.~\ref{Fig:3}(d-i) and Supplemental Movie~\red{S2}). 

In regime II, with increasing $\Wen$, there is a sharp increase in the amplitude $\tilde{F}_2$ of the second peak. A striking feature of this regime is that the magnitude of the second peak force exceeds that of the first one, $\tilde{F}_2 > \tilde{F}_1$, see Fig.~\ref{Fig:3}(c) which  illustrates the case with the highest second peak force ($\tilde{F}_2 = 2.98$, occurring  for $\Wen = 9$, Supplemental Movie~\red{S3}). The large force amplitude in this regime correlates to the formation of an ultra-thin and high-velocity singular Worthington jet \cite{Bartolo2006Singular}. Here, the Worthington jet is most pronounced as it results from the collapse of an air-cavity as well as the converging capillary waves (see insets of Fig.~\ref{Fig:3}(c) and~\ref{Fig:3}(d-ii - d-iii)). It is reminiscent of the hydrodynamic singularity that accompanies the bursting of bubbles at liquid-gas interfaces \cite{woodcock1953giant, sanjay_lohse_jalaal_2021}. Outside regime II such bubbles do not form, see Fig.~\ref{Fig:3}(d-iv - d-vi). Consistent with this view, the case with maximum peak force [$\Wen = 9$, Fig.~\ref{Fig:3}(c)] entrains the largest bubble. Another characteristic feature of this converging flow is that, despite having a small Ohnesorge number ($= 0.0025$) that is often associated with inviscid potential flow inside the drop \cite{molavcek2012quasi}, it still shows high rates of local viscous dissipation near the axis of symmetry [Fig~\ref{Fig:3}(c) insets and~\ref{Fig:3}(d-ii - d-iii)], due to the singular character of the flow (see Supplemental Movie~\red{S4}).  

When $\Wen$ is further increased, we (locally) find $\tilde{F}_2 \sim \Wen^{-1}$ in the transition regime ($12.6 < \Wen < 30$), followed by $\tilde{F}_2 \sim \Wen^0$ in the inertial regime III ($30 < \Wen < 100$). Specifically, by employing best fits, we obtain
\begin{equation}
	\tilde F_2 = \frac{F_2}{\rho_d V_0^2 D_0^2} \approx
	\begin{cases}
		11\Wen^{-1} &(12.6 < \Wen < 30),\\
		0.37 &(30 < \Wen < 100).
	\end{cases}
	\label{eq_1}
\end{equation}

We will now rationalize this experimentally and numerically observed scaling behavior of the amplitude $F_2$ of the second peak using scaling arguments. As already mentioned, Fig.~\ref{Fig:3}(d) shows that the second peak in the force at $t_2$ coincides with an upwards jet, which has typical velocity $v_j$ (see Supplemental Material \S~\red{IV} for calculation details) and typical diameter $d_j$, Fig.~\ref{Fig:3}(e). Fig.~\ref{Fig:3}(d) also illustrates strong radially symmetric flow focusing due to the retracting drop in regimes II and III. We define the recoiling velocity of the drop at time $t_2$ as $v_2$, the droplet height at that moment as $h_2$, and the droplet diameter at that moment as $D_2 = D(t_2)$, see again Fig.~\ref{Fig:3}(e). Note that regime II also includes stronger converging capillary waves and the collapsing air cavity  [Fig.~\ref{Fig:3}(c) insets and Fig.~\ref{Fig:3}(d-ii - d-iii)]. The presence of the substrate breaks the symmetry in vertical direction, directing the flow into the Worthington jet. Using continuity and balancing the volume flux at this instant $t_2$, we obtain $v_2  D_2 h_2 \sim v_jd_j^2$. Of course, $D_2$ and $h_2$ are also related by volume conservation. Assuming a pancake-type shape at $t_2$, we obtain $D_2^2h_2 \sim D_0^3$ \cite{Wildeman2016} and therefore, $v_jd_j^2 \sim v_2 D_0^3/D_2$. As the drop retracts, the velocity of the flow field far away from the jet is parallel to the base [Fig.~\ref{Fig:3}(d)]. So, the occurrence and strength of the second peak $F_2$ is mainly a result of the flow opposite to the vertical Worthington jet [Fig.~\ref{Fig:3}(d-iv - d-vi)], which naturally leads to $F_2 \sim \rho_d v_j^2d_j^2$ (momentum flux balance in the vertical direction). Combining the above arguments, we get $F_2 \sim \rho_d v_jv_2 D_0^3/D_2$ which can be non-dimensionalized with the inertial pressure force $\rho_d V_0^2D_0^2$ to obtain
\begin{equation}
	\tilde{F}_2 = \frac{F_2}{\rho_d V_0^2D_0^2} \sim \frac{\tilde{v}_j\tilde{v}_2}{\tilde D_2},
	\label{Eqn::TheoryForce2}
\end{equation}
\noindent
where, $\tilde{v}_j = v_j/V_0$, $\tilde{v}_2 = v_2/V_0$, and $\tilde D_2 = D_2/D_0$ are the dimensionless jet velocity, drop retraction velocity, and spreading diameter, respectively, all at $t_2$.  

Fig.~\ref{Fig:3}(f) compares the amplitude of the second peak as obtained from the experiments and simulations with the theoretical prediction of Eq.~(\ref{Eqn::TheoryForce2}) (see Supplemental Material \S~\red{IV}). Indeed, this scaling relation reasonably well describes the transitional regime II-III and regime III data. Obviously, in regime I, the theoretical prediction is invalid because the hypothesis of flow focusing breaks down, and capillary oscillations dominate the flow, with no Worthington jet occurring [Fig.~\ref{Fig:3}(d-i)]. Further, Eq.~(\ref{Eqn::TheoryForce2}) over-predicts the forces in regime II because efficient capillary waves focusing and air cavity collapse lead to extremely high-velocity singular jets. The entrained air bubble also shields momentum transfer from the singular Worthington jet to the substrate [insets of Fig~\ref{Fig:3}(c)]. 

We finally come to the very large impact velocities of regime IV. Then, when $\Wen \gtrsim 100$, in the experiments splashing occurs \cite{Josserand2016}, see Supplemental Movie~\red{S6}. At such high $\Wen$, the surrounding gas atmosphere destabilizes the rim \cite{Eggers2010, riboux2014experiments}. Therefore, in regime IV, kinetic and surface energies are lost due to the formation of satellite droplets, resulting in diminishing $\tilde{F}_2$ in the experiments [Fig.~\ref{Fig:3}(b)]. In contrast, for our axisymmetric (by definition) simulations, the above-mentioned azimuthal instability is absent \cite{Eggers2010} and the plateau $\tilde{F}_2 \approx 0.37$ continues in this regime. Consequently, Eq.~(\ref{Eqn::TheoryForce2}) holds only for the simulations in regime IV [Fig.~\ref{Fig:3}(f)], and not for the experiments. Further analysis of the experimentally observed fragmentation scenario is beyond the scope of the present work. For future work, we suggest that one could also experimentally probe $F_2$ in this regime by suppressing the azimuthal instability (for instance, by reducing the atmospheric pressure \cite{xu2005drop}). 
  
\noindent
{\it Conclusions and outlook:} 
In this Letter, we have experimentally obtained the normal force profile $F(t)$ of water drops impacting superhydrophobic surfaces. To elucidate the physics and study the internal flow, we have also used direct numerical simulations, which perfectly match the experimental results without any fitting parameter. In the force profile $F(t)$, we identified two prominent peaks. The first peak arises from an inertial shock following the impact of the impacting drop onto the immobile substrate. The hitherto unknown second peak occurs before the drop rebounds. The variation of the amplitude of this peak with Weber number divides the parameter space into four regimes, namely the capillary, singular jet, inertial, and splashing regime. This peak in the force occurs due to the momentum balance when the Worthington jet is created by flow focusing, owing either to capillary waves (singular jet regime) or drop retraction (inertial regime). Surprisingly, even a low Weber number impact (singular jet regime) can lead to a highly enhanced 2nd peak in the force profile, remarkably even larger than the first one, 
triggered by the singularity occurring at the collapse of an air cavity. Lastly, we have derived
scaling relations for these peak forces. 

The esthetic beauty of our results on the drop impact dynamics on a non-wetting surface and the forces associated with it lies in the combination of the simplicity and daily-life character of the experiment and the observed rich and surprising  phenomenology. The achieved insight is technologically relevant to develop countermeasures to the failure of superhydrophobic materials (for e.g., by avoiding the regime $5.3 < \Wen < 12.6$ or reducing the spacing of the textures \cite{Lafuma2003}). Interesting extensions of our work include the study of impact forces of viscous drops (i.e., drops with $\Ohn < 1$), which will show quite different scaling behavior \cite{Jha2020}, and of Leidenfrost drops \cite{quere2013}.

\begin{acknowledgments}
This work received financial support from the National Natural Science Foundation of China (no. 11872227, 11902179, 11632009, 11921002), and support from Tsinghua University (no. 53330100321). We also acknowledge funding from the ERC Advanced Grant DDD under Grant No. 740479. The numerical simulations were carried out on the national e-infrastructure of SURFsara, a subsidiary of SURF cooperation, the collaborative ICT organization for Dutch education and research. The authors are grateful to Marie-Jean Thoraval, Uddalok Sen, and Pierre Chantelot for the stimulating discussions. B.Z. thanks Maosheng Chai for SEM testing supports.
\end{acknowledgments}

\bibliography{ms}

\end{document}



\title{Supplemental Material:\\ Impact forces of water drops falling on superhydrophobic surfaces}

\author{Bin Zhang}

\affiliation{%
	Department of Engineering Mechanics, AML, Tsinghua University, Beijing 100084, China
}%
\author{Vatsal Sanjay}
\affiliation{
	Physics of Fluids Group, Max Planck Center Twente for Complex Fluid Dynamics, MESA+ Institute, and J. M. Burgers Center for Fluid Dynamics, University of Twente, P.O. Box 217, 7500AE Enschede, Netherlands
}
\author{Songlin Shi}
\affiliation{%
	Department of Engineering Mechanics, AML, Tsinghua University, Beijing 100084, China
}%
\author{Yinggang Zhao}
\affiliation{%
	Department of Engineering Mechanics, AML, Tsinghua University, Beijing 100084, China
}%
\author{Cunjing Lv}
\email{cunjinglv@tsinghua.edu.cn}
\affiliation{%
	Department of Engineering Mechanics, AML, Tsinghua University, Beijing 100084, China
}%
\author{Xi-Qiao Feng}
\affiliation{%
	Department of Engineering Mechanics, AML, Tsinghua University, Beijing 100084, China
}%
\author{Detlef Lohse}
\affiliation{	
	Physics of Fluids Group, Max Planck Center Twente for Complex Fluid Dynamics, MESA+ Institute, and J. M. Burgers Center for Fluid Dynamics, University of Twente, P.O. Box 217, 7500AE Enschede, Netherlands
}
\affiliation{
	Max Planck Institute for Dynamics and Self-Organisation, Am Fassberg 17, 37077 G{\"o}ttingen, Germany
}
\date{\today}

\maketitle

\tableofcontents

\clearpage
\section{Methods}
\begin{figure}
	\includegraphics[width=\linewidth]{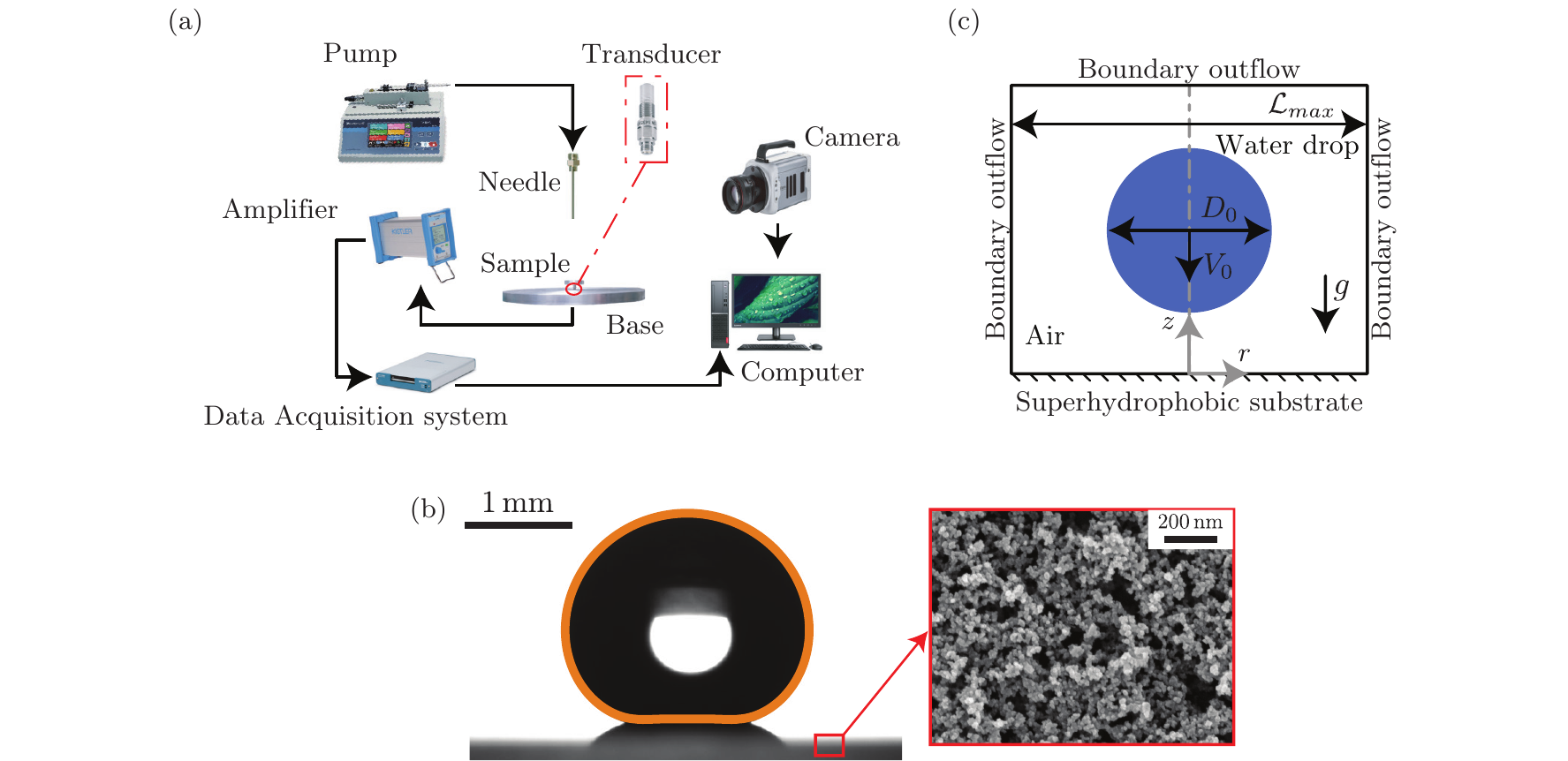}
	\caption{(a) Experimental setup, consisting of four main units: a drop generation unit, a superhydrophobic substrate, a force measurement unit, and a high-speed photography unit. Relevant appurtenant devices are shown. Also see Fig.~\red{1}(a) of the main manuscript. (b) Snapshot of a water drop sitting on a superhydrophobic substrate with overlaid orange boundary from simulation. The inset shows the scanning electron microscope (SEM) image of the superhydrophobic surface covered by hydrophobic nanoparticles. (c) Simulation domain with the appropriate boundary conditions. The boundaries are kept far away from the drop to avoid feedback ($\mathcal{L}_{max} \gg D_0$).}
	\label{Fig_domain}
\end{figure}

\subsection{Experimental method}
\label{sec:ExpMethods}
Fig.~\ref{Fig_domain}a illustrates the experimental setup consists of four main units: a drop generation unit, a superhydrophobic substrate, a force measurement unit and a high-speed photography unit (also see Fig.~\red{1}(a) of the main manuscript). 
\subsubsection{Drop generation unit}
The drop generation unit is used to create drops of different sizes and impact velocities independently (initial drop diameter $D_0$ ($2.05\,\si{\milli\meter} \le D_0 \le 2.76\,\si{\milli\meter}$) and the impact velocity $V_0$ ($0.38\,\si{\meter}/\si{\second} \le V_0 \le 2.96\,\si{\meter}/\si{\second}$)). Deionized water drops are created by employing suspended needles that are connected to a syringe pump. To suppress disturbances, the drop is created at a very smooth flow rate, i.e., $0.5\,\si{\milli\liter}/\si{\minute}$. The drop diameter is varied by employing needles of different sizes, and the impact velocity is controlled by changing the distance between the needle and the sample by employing a vertical translation stage.

\subsubsection{Superhydrophobic substrate}\label{sec::substrate}
A water drop impacts a quartz plate whose upper surface is coated with silanized silica nanobeads with diameter of $20\,\si{\nano\meter}$ (Glaco Mirror Coat Zero; Soft99) \cite{Li2017, Gauthier2015}  to attain superhydrophobicity [Fig.~\ref{Fig_domain}(b)]. The advancing and receding contact angles of water drops are $167\,\pm\,2^\circ$ and $154\,\pm\,2^\circ$, respectively. 

\begin{figure}
	\includegraphics[width=\linewidth]{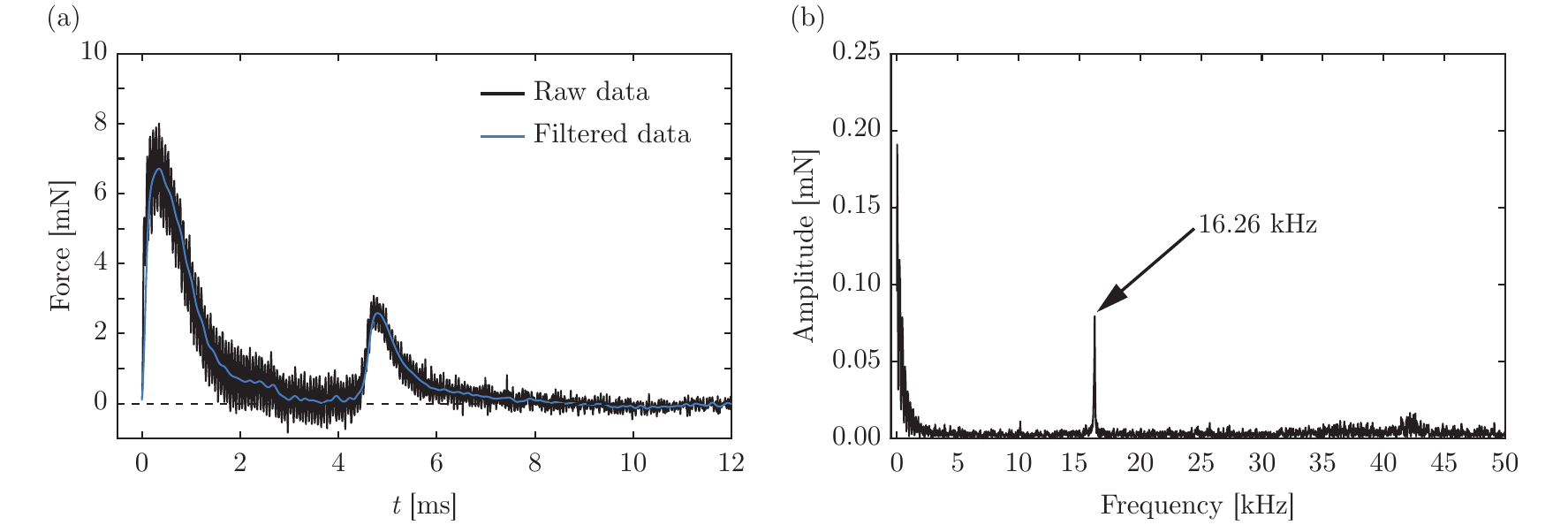}
	\caption{(a) Raw force data (black curve) without filtering and filtered force profile (blue curve) for an impinging drop on the superhydrophobic surface for $D_0 = 2.05\,\si{\milli\meter}$ and $V_0 = 1.39\,\si{\meter}/\si{\second}$ giving $\Wen = 54$. (b) Frequency spectrum of the raw force data.}
	\label{Fig_Filtering}
\end{figure}

\subsubsection{Force measurement unit}
\label{sec:Sample}
The core apparatuses of the force measurement unit include a sample, a high-precision transducer, a charge amplifier and a data acquisition system. The lower surface of the quartz plate is glued to an aluminum base, which is screwed vertically into a high-precision piezoelectric force transducer (Kistler 9215A) with a resolution of $0.5\,\si{\milli\newton}$. If the magnitude of the normal reaction force between the drop and the substrate is smaller than this resolution ($0.5\,\si{\milli\newton}$) of our force transducer and cannot be isolated from noise. The impact force between the drop and the sample is measured by the collection of the charge generated by the transducer, and the charge is immediately converted into a voltage by employing an amplifier (Kistler 5018A). Comparing to the weak charge signal that is sensitive to the environment noise (such as triboelectricity due to the movement of the cable, or magnetic fields in the environment, etc.), the amplified voltage signal is more robust for transmitting and processing. After that, the amplified analog signals are converted into the digital signals by employing a data acquisition system (NI USB-6361 driven by Labview) at a sampling rate of 100 kHz. Finally, the unit of the measured signal is changed from Voltage ($\si{\volt}$) to Newton ($\si{\newton}$) via the calibration coefficient of the force transducer. 

In our experiment, high-frequency vibrations of the experimental setup are inevitably induced by the drop impact, and they will superimpose on the temporal evolution of the impact force. As shown in Fig.~\ref{Fig_Filtering}(a), the raw force data of a drop impacting on the superhydrophobic surface ($D_0 = 2.05\,\si{\milli\meter}, V_0 = 1.39\,\si{\meter}/\si{\second}$ giving $\Wen = 54$) has two peaks with obvious oscillations along the curve. The frequency spectrum of the raw data shown in Fig.~\ref{Fig_Filtering}(b) helps us identifying the frequency relating to the vibration of the measurement system (i.e., $16.26\,\si{\kilo\hertz}$). To obtain the signal of the impact force, the oscillations are spectrally removed by employing a low pass filter with a cut-off frequency of $5\,\si{\kilo\hertz}$. The similar filter method was also employed in \cite{Li2014, Zhang2017, Gordillo2018, Mitchell2019}. Finally, we get the impact force curve (i.e., blue curve) as shown in Fig.~\ref{Fig_Filtering}(a).

\subsubsection{High-speed photography}
Lastly, the fast force sensing technique described above is synchronized with the high-speed photography unit containing a high-speed camera (Photron Fastcam Nova S12) and a micro Nikkor 105 mm f/2.8 imaging lens. To realize a synchronization of the evolutions of the drop morphology and the transient force, the high-speed camera is triggered by the data acquisition system when the impact force is larger than $1\,\si{\milli\newton}$. A LED light (CLL-1600TDX) of adjustable output power is used to illuminate the scene of the impingement. We efficiently record the drop impact phenomenon at $10,000$ fps with a shutter speed $1/20,000\,\si{\second}$.

\subsection{Numerical method}\label{sec:NumMethods}
This section elucidates the direct numerical simulation framework used to study the drop impact process [Fig.~\ref{Fig_domain}(c)] using the free software program, \emph{Basilisk C} \cite{popinet-basilisk}. 

\subsubsection{Governing equations}
For an incompressible flow, the mass conservation requires the velocity field to be divergence-free $\left(\boldsymbol{\nabla\cdot v} = 0\right)$.  Furthermore, the momentum conservation  reads 
\begin{align}
	\label{Eqn::NS}
	\rho\left(\frac{\partial}{\partial t} + \boldsymbol{v\cdot\nabla}\right)\boldsymbol{v} = -\boldsymbol{\nabla} p^{\prime} - [\rho]\left(\boldsymbol{g}\cdot\boldsymbol{z}\right)\hat{\boldsymbol{n}}\delta_s + \boldsymbol{\nabla\cdot}\left(2\mu\boldsymbol{\mathcal{D}}\right) + \boldsymbol{f}_\gamma.
\end{align}
\noindent Here, the bracketed term on the left hand side is the material derivative. On the right hand side, the deformation tensor, $\boldsymbol{\mathcal{D}}$ is the symmetric part of the velocity gradient tensor $\left(\boldsymbol{\mathcal{D}} = \left(\nabla\boldsymbol{v} + \left(\nabla\boldsymbol{v}\right)^{\text{T}}\right)/2\right)$. Further, $p^{\prime}$ denotes reduced pressure field, $p' = p - \rho\boldsymbol{g}\cdot\boldsymbol{z}$, where, $p$ and $\rho\boldsymbol{g}\cdot\boldsymbol{z}$ represent the dynamic and the hydrostatic pressures, respectively, with $\boldsymbol{g}$ and $\boldsymbol{z} = z\hat{\boldsymbol{z}}$ representing the gravitational acceleration vector and the vertical coordinate vector, respectively ($z$ is the distance away from the superhydrophobic substrate and $\hat{\boldsymbol{z}}$ is a unit vector, see Fig.~\ref{Fig_domain}(c)). Using this reduced pressure approach ensures an exact hydrostatic balance as described in \citet{popinet2018numerical}, and \citet{basiliskPopinet3}. Note that this formulation requires an additional singular body force $\left([\rho]\left(\boldsymbol{g}\cdot\boldsymbol{z}\right)\hat{\boldsymbol{n}}\delta_s\right)$ at the interface. Here, $[\rho]$ is the density jump across the interface, $\hat{\boldsymbol{n}}$ is the interfacial normal vector, $\hat{\boldsymbol{n}} = \boldsymbol{\nabla}H/\|\boldsymbol{\nabla}H\|$, and $\delta_s$ is the \emph{Dirac}-delta function, $\delta_s = \|\boldsymbol{\nabla}H\|$, where $H$ is the Heaviside function. Consequently, $\delta_s$ is non-zero only at the liquid-air interface and has units of $1/\text{length}$ (for detailed derivation, see appendix B of \citet{tryggvason2011direct}). Furthermore, we employ one-fluid approximation \cite{prosperetti2009computational, tryggvason2011direct} to solve these equations employing volume of fluid (VoF) method for interface tracking, whereby the Heaviside function can be approximated by the VoF marker function $\Psi$ ($\Psi = 1$ inside the liquid drop, and $\Psi = 0$ in the air). Subsequently, the material properties (such as density $\rho$ and viscosity $\mu$) change depending on which fluid is present at a given spatial location,

\begin{align}
	\label{Eqn::density}
	\rho &= \Psi\rho_d + \left(1-\Psi\right)\rho_{a},\\
	\label{Eqn::Oh}
	\mu &= \Psi\mu_d + \left(1-\Psi\right)\mu_{a},
\end{align}

\noindent where, the subscripts $d$ and $a$ denote drop and air, respectively. The VoF marker function ($\Psi$) follows the advection equation \cite{prosperetti2009computational, tryggvason2011direct}, 

\begin{equation}
	\label{Eqn::Vof2}
	\left(\frac{\partial}{\partial t} + \boldsymbol{v\cdot\nabla}\right)\Psi = 0.
\end{equation}

Lastly, a singular body force $\boldsymbol{f}_\gamma$ is applied at the interfaces to respect the dynamic boundary condition. The approximate forms of this force follows \cite{brackbill1992continuum}

\begin{equation}\label{Eqn::SurfaceTension}
	\boldsymbol{f}_\gamma = \gamma\kappa\delta_{s}\hat{\boldsymbol{n}} \approx \gamma\kappa\boldsymbol{\nabla}\Psi.
\end{equation}

\noindent Here, $\gamma$ is the drop-air surface tension coefficient and $\kappa$ is the curvature calculated using the height-function method.  During the simulations, the maximum time-step needs to be less than the oscillation period of the smallest wave-length capillary wave because the surface-tension scheme is explicit in time \citep{popinet2009accurate, basiliskPopinet2}.

In our simulations, ideal superhydrophobicity is maintained by assuming that a thin air layer is present between the drop and the substrate \cite{ramirez2020lifting}. The normal force on this substrate can be calculated using \cite{landau2013course}

\begin{equation}
	\label{Eqn::force}
	\boldsymbol{F}(t) = \int_\mathcal{A} \left(\left(p-p_0\right)\left(\boldsymbol{I}\boldsymbol{\cdot}\hat{\boldsymbol{z}}\right) - 2\mu_a\left(\boldsymbol{\mathcal{D}}\boldsymbol{\cdot}\hat{\boldsymbol{z}}\right)\right) \mathrm{d}\mathcal{A},
\end{equation}

\noindent where, $p$ and $p_0$ are the dynamic pressure distribution at the substrate and the ambient pressure, respectively. Here, $\boldsymbol{I}$ is the second-order identity tensor. Further, $\hat{\boldsymbol{z}}$ is the unit vector normal to the substrate [Fig.~\ref{Fig_domain}(c)] and $\mathcal{A}$ represents substrate's area. Note that the contribution from the second term on the right-hand side of Eq.~(\ref{Eqn::force}) is the normal viscous force due to the air layer between the drop and the substrate and is negligible as compared to the pressure integral. Therefore, we can calculate the normal impact force simply by integrating the pressure field at the substrate, 

\begin{equation}
	\label{Eqn::force2}
	\boldsymbol{F}(t) = F(t) \hat{\boldsymbol{z}} = \left(\int_\mathcal{A} \left(p-p_0\right)\mathrm{d}\mathcal{A}\right)\hat{\boldsymbol{z}}.
\end{equation}

Despite a low viscosity associated with the water drops, the viscous dissipation can still be significant in some cases, especially during flow focusing and capillary waves resonance (see Fig.~\red{3} of the main text). To identify these regions of high viscous dissipation, we also measure the rate of viscous dissipation per unit volume, given by \cite{landau2013course}

\begin{equation}
	\dot{E}_d^{\text{local}} = 2\mu \left({\boldsymbol{\mathcal{D}}:\boldsymbol{\mathcal{D}}}\right),
	\label{Eqn::dissipation1}
\end{equation}

\noindent which on non-dimensionalization using the drop diameter ($D_0$), density ($\rho_d$), and impact velocity ($V_0$) gives

\begin{equation}
	\dot{\tilde{E}}_d^{\text{local}} \equiv \frac{\dot{E}_d^{\text{local}}}{\rho_dV_0^3/D_0} = \frac{2}{Re}\left(\Psi + \frac{\mu_a}{\mu_d}\left(1-\Psi\right)\right)\left(\boldsymbol{\tilde{\mathcal{D}}:\tilde{\mathcal{D}}}\right),
	\label{Eqn::dissipation2}
\end{equation}

\noindent where, the Reynolds number ($Re = \rho_d V_0D_0/\mu_d$) is the ratio of inertial to viscous stresses, and $\boldsymbol{\tilde{\mathcal{D}}} = \boldsymbol{\mathcal{D}}/(V_0/D_0)$. 

\subsubsection{Relevant dimensionless numbers}
In the experiments, the initial drop diameter $D_0$ ($2.05\,\si{\milli\meter} \le D_0 \le 2.76\,\si{\milli\meter}$) and the impact velocity $V_0$ ($0.38\,\si{\meter}/\si{\second} \le V_0 \le 2.96\,\si{\meter}/\si{\second}$) are independently controlled. The drop material properties are kept constant (density $\rho_d = 998\,\si{\kilogram}/\si{\meter}^{3}$, surface tension coefficient $\gamma = 73\,\si{\milli\newton}/\si{\meter}$, and dynamic viscosity $\mu_d = 1.0\,\si{\milli\pascal}\si{\second}$). As a result, we identify the following dimensionless numbers,

\begin{align}
	\Wen &= \frac{\rho_dV_0^2D_0}{\gamma} \\
	\Ohn &= \frac{\mu_d}{\sqrt{\rho_d\gamma D_0}}\\
	Bo &= \frac{\rho_dgD_0^2}{\gamma}
\end{align}

where, $\Wen$ is the impact Weber number which is a ratio of the inertial to capillary pressures. The Ohnesorge number ($\Ohn$) is the ratio between the inertial-capillary to the inertial-viscous time scales and is kept constant at $0.0025$ to mimic $2\,\si{\milli\meter}$ diameter water drops. Furthermore, the Bond number ($Bo$) is the ratio of the gravitational to the capillary pressure, which is also fixed at $0.5$ for the same reason. Lastly, to minimize the influence of the surrounding medium, $\rho_a/\rho_d$ and $\mu_a/\mu_d$ are fixed at $10^{-3}$ and $3 \times 10^{-3}$, respectively. 

\subsubsection{Domain description}
Fig.~\ref{Fig_domain}(c) represents the axi-symmetric computational domain where $r = 0$ denotes the axis of symmetry. A no-slip and non-penetrable boundary condition is applied on the substrate along with zero pressure gradient. Here, we also use $\Psi = 0$ to maintain a thin air layer between the drop and the substrate. Physically, it implies that the minimum thickness of this air layer is $\Delta$ throughout the whole simulation duration (where $\Delta$ is the minimum grid size). Further, boundary outflow is applied at the top and side boundaries (tangential stresses, normal velocity gradient, and ambient pressure are set to zero). 

Furthermore, the domain boundaries are far enough not to influence the drop impact process ($\mathcal{L}_{\text{max}} \gg D_0$). \emph{Basilisk C} \cite{popinet-basilisk} also allows for Adaptive Mesh Refinement (AMR) with maximum refinement in the regions of high velocity gradients and at the drop-air interface. We also undertook a mesh independence study to ensure that the results are independent of this mesh resolution. We use a minimum grid size $\Delta = D_0/1024$ for this study. Note that the cases in regime II (singular jet) requires a refinement of $\Delta \approx D_0/4098$ near the axis of symmetry. The simulation source codes, as well as the post-processing codes used in the numerical simulations, are permanently available at the \emph{Github} repository \cite{basiliskVatsal}. 

\subsubsection{Calculating the time of contact}

\begin{figure}
	\centering
	\includegraphics[width=\linewidth]{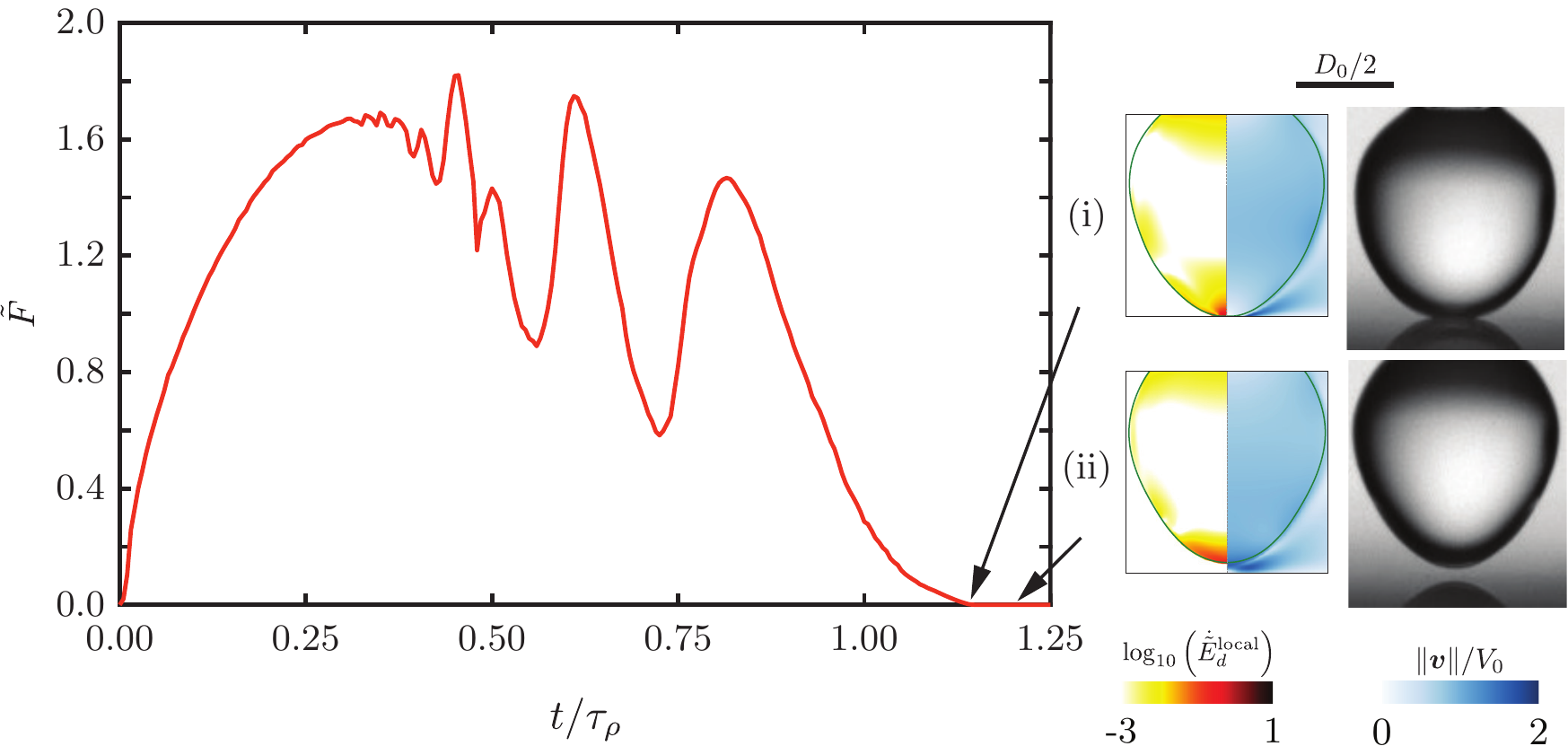}
	\caption{Temporal variation of the normal contact force for $\Wen = 2$. The insets show two key instances: (i) time $t_3 = 1.14\tau_\rho$ when $F$ vanishes which marks the contact time of the drop at the substrate and (ii) detachment time ($1.2\tau_\rho$) as seen from the side view image. Also see supplemental movie~\red{S2}.}
	\label{Fig_TakeOff}
\end{figure}

In the main text, we emphasize that the time instant $t_3$ at which the normal contact force between the drop and the substrate vanishes is a much better estimate for the drop contact time as compared to the one observed at complete detachment from side view images. This observation is consistent with the literature \cite{Bouwhuis2012, van2012direct, lee2020drop,  chantelot_lohse_2021} and is elucidated in Fig.~\ref{Fig_TakeOff} for $\Wen = 2$. In this case, $t_3 = 1.14\tau_\rho$ whereas the side view images show complete detachment at $t = 1.2\tau_\rho$. The effect is further enhanced for higher $\Wen$, see for comparisons, supplemental movies~\red{S1} ($\Wen = 40$) and~\red{S2} ($\Wen = 2$).

\section{Comparisons of impact force between superhydrophobic and hydrophilic surfaces}\label{sec:PhobicPhilic}
\begin{figure}
	\includegraphics[width=\linewidth]{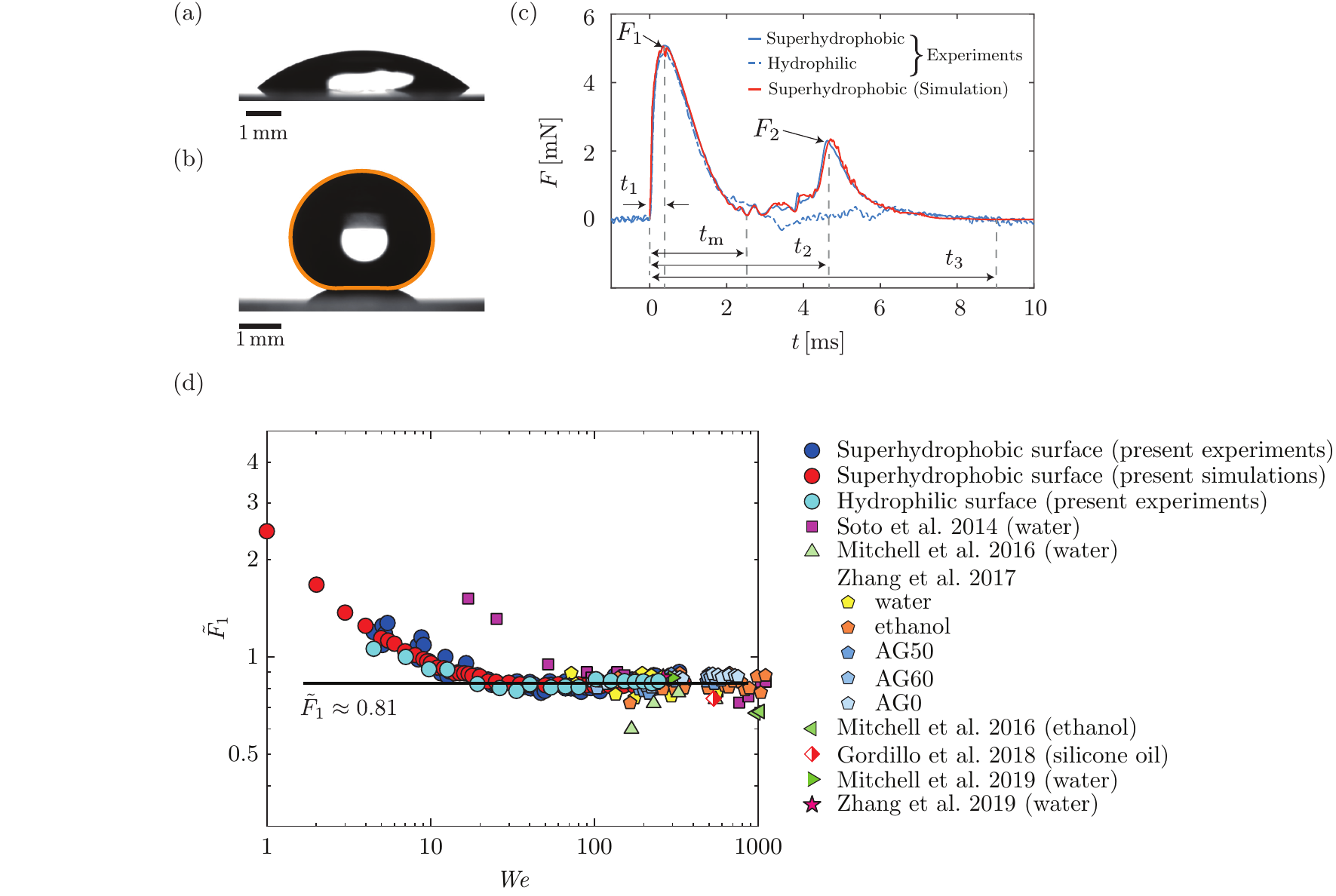}
	\caption{Wettability of the surfaces and the corresponding transient force profiles. Water drops depositing on (a) hydrophilic and (b) superhydrophobic surfaces with apparent contact angles of $40\,\pm\,4^\circ$ and $165\,\pm\,1^\circ$, respectively. For the superhydrophobic case, the drop boundary from simulation is overlaid in orange. (c) Transient force profiles on the hydrophilic and superhydrophobic surfaces. The initial diameter of the drops is $2.05\,\si{\milli\meter}$, and the impact velocity is $1.20\,\si{\meter}/\si{\second}$, corresponding to $\Wen = 40$. (d) Variation of the first dimensionless peak force $\tilde{F}_1$ as a function of $\Wen$.  Also see Fig.~\red{1(c)} and~\red{3(c)} of the main text.}
	\label{Fig_phobicphilic}
\end{figure}

To differentiate between impact forces on superhydrophobic surfaces to that of hydrophilic ones \cite{Li2014, Soto2014, Philippi2016, Zhang2017, Gordillo2018, Mitchell2019, Zhang2019}, we carry out test impacts on hydrophilic surfaces. The hydrophilic sample is a quartz plate, cleaned by surfactant, deionized water, alcohol and deionized water in sequence before the experiment. The advancing and receding contact angles of the deionized water drops on the quartz surface are $47\,\pm\,2^\circ$ and $13\,\pm\,2^\circ$, respectively [Fig.~\ref{Fig_phobicphilic}a]. The superhydrophobic surface is a Glaco-coated quartz plate \cite{Li2017, Gauthier2015} as described in \S~\ref{sec::substrate}, on which the advancing and receding contact angles are $167\,\pm\,2^\circ$ and $154\,\pm\,2^\circ$, respectively [Fig.~\ref{Fig_phobicphilic}(b)]. 

Fig.~\ref{Fig_phobicphilic}(c) compares the impact on superhydrophobic and hydrophilic substrates for impact corresponding to $\Wen = 40.4$ ($D_0 = 2.05\,\si{\milli\meter}$ and $V_0 = 1.20\,\si{\meter}/\si{\second}$). The comparison shows that in the spreading stage ($0 < t < 2\,\si{\milli\second}$), the transient force profiles overlap. In the time span $2\,\si{\milli\second} < t < 9\,\si{\milli\second}$, the transient force profile of the drop impact on the hydrophilic surface only has slight fluctuations around zero. In contrast to the hydrophilic one, there is an obvious peak force (i.e. $F_2$, corresponding to $t \approx 4.63\,\si{\milli\second}$) in the retraction stage of the drop impact on the superhydrophobic surface. 

Furthermore, the impact force $F_1$ on the superhydrophobic surface is equal to the maximum impact force on the hydrophilic surface. To obtain a comprehensive understanding, we extracted experimental data (the maximum impact force) from previous literature performed on hydrophilic surfaces \cite{Soto2014, mitchell2016experimental, Zhang2017, Gordillo2018, Mitchell2019, Zhang2019}. Moreover, we carried out experiments on hydrophilic quartz surfaces with an apparent contact angle of $40 \pm 4\si{\degree}$. Then, as shown in Fig.~\ref{Fig_phobicphilic}(d), we make a comparison of $F_1$ between previous work (on hydrophilic surfaces) and our work (on both hydrophilic and superhydrophobic surfaces). As shown in Fig.~\ref{Fig_phobicphilic}(d), the data on both superhydrophobic and hydrophilic surfaces in our study are consistent with each other. Furthermore, when $We > 30$, the data in the present work and previous literature are consistent with each other. Therefore, $F_1$ only depends on the Weber number, rather than the wettability of the surface.

\section{Some notes on the different regimes of drop impact}

\subsection{Regime II: singular Worthington jet}
\begin{figure}
\includegraphics[width=\linewidth]{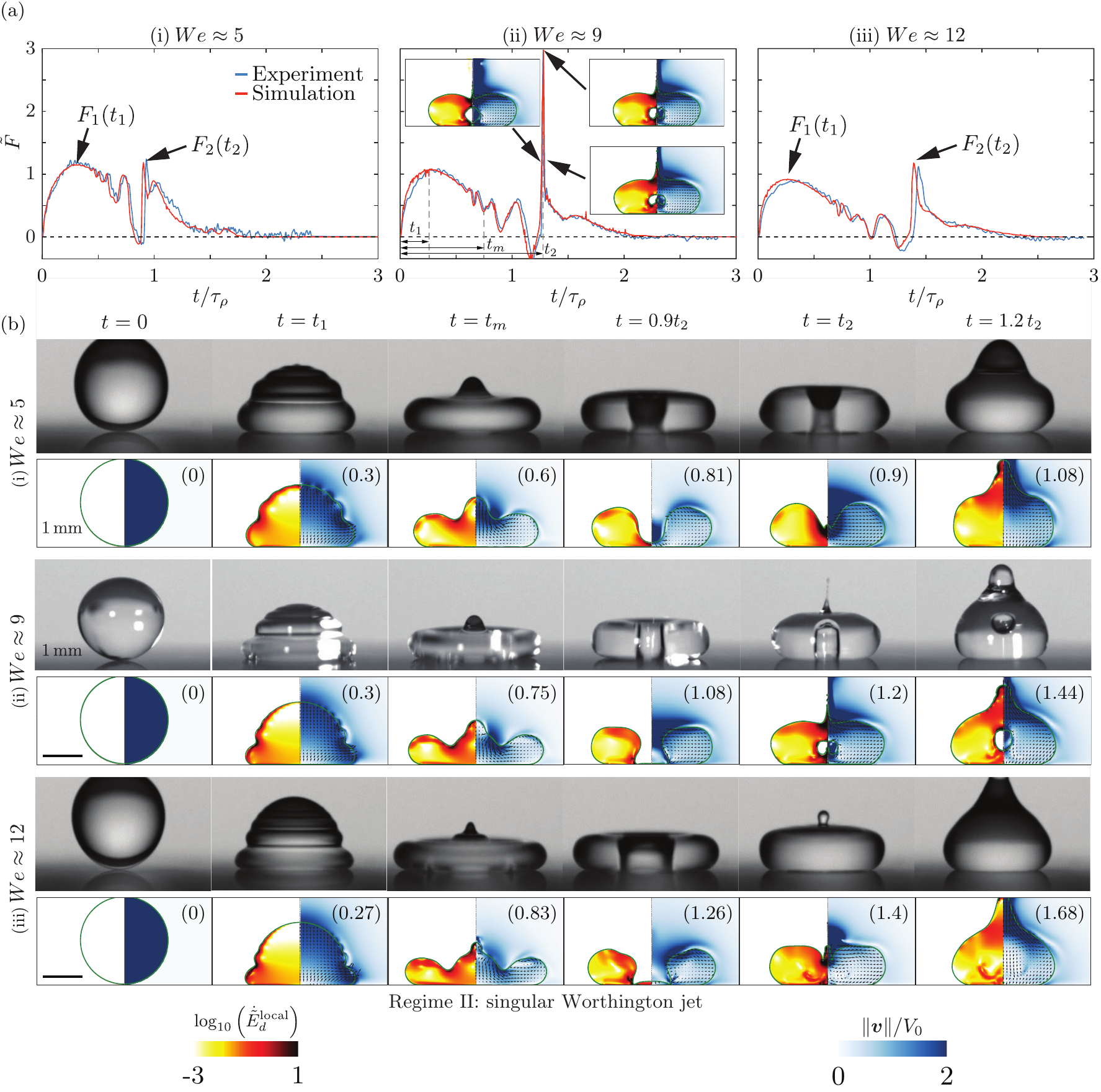}
\caption{Drop impact on the superhydrophobic surface in the singular Worthington jet regime. (a) Evolution of the normal force $\tilde{F}(t) = F(t)/\rho_dV_0^2D_0^2$ of an impacting drop for the case with high $\tilde{F}_2$. Insets show the drop morphology and flow anatomy close to the capillary resonance that leads to a hydrodynamic singularity. Note the outstanding agreement between the experimental (blue line) and the numerical (red line) results, including the various wiggles in the curve, which originate from capillary oscillations. (b) Impacts at different Weber numbers in regime II showing experimental snapshots in the top row and numerical simulations in the bottom}. In both panels (a) and (b), $\Wen \approx$ (i) $5$, (ii) $9$, and (iii) $12$. The numbers in the top right of each numerical snapshot mentions the dimensionless time $t/\tau_\rho$. The left part of each numerical snapshot shows the dimensionless local viscous dissipation rates on a $\log_{10}$ scale and the right part shows the velocity field magnitude normalized with the impact velocity. The black velocity vectors are plotted in the center of mass reference frame of the drop to clearly elucidate the internal flow. Also see Supplemental Movies~\red{S3} and~\red{S4}.
\label{Fig_RegimeII}
\end{figure}

In the main text, we discussed several features of regime II. In this section, we further elucidate this regime using three representative cases in Fig.~\ref{Fig_RegimeII}. We replot the data for $\Wen = 9$ which shows the maximum force amplitude and choose $\Wen = 5$ and $\Wen = 12$ near the boundaries of regime II (also see Supplemental Movie~\red{S4}). Here, we look at both the transient force profile [Fig.~\ref{Fig_RegimeII}(a)] and the anatomy of flow inside the drops [Fig.~\ref{Fig_RegimeII}(b)]. The transient force profiles show similar features for these three different Weber numbers. After the impact at $t = 0$, there is a sharp increase in the force which reaches the maximum at $t = t_1$ [Fig~\ref{Fig_RegimeII}(a)]. As the drops spread further, their morphology [Fig.~\ref{Fig_RegimeII}(b)] feature distinct pyramidal structures owing to the capillary waves \cite{renardy2003pyramidal} that manifest as oscillations in the temporal evolution of the forces. Then, the drop spreads to a maximum radial extent at $t = t_m$ followed by the retraction phase as the surface tension pulls the drop radially inwards, further enhancing the capillary waves. These traveling capillary waves interact to form an air-cavity, for instance, see $t = 0.9t_2$ in Fig.~\ref{Fig_RegimeII}b. The cavity collapses to create high-velocity singular Worthington jets. Subsequently, bubble is entrained (see $\Wen = 9$, Fig.~\ref{Fig_RegimeII}(b-ii)). Comparing the force profile for $\Wen = 9$ [Fig.~\ref{Fig_RegimeII}(a-ii)] with that of $\Wen = 5$ [Fig.~\ref{Fig_RegimeII}(a-i)] and $\Wen = 12$ [Fig.~\ref{Fig_RegimeII}(a-iii)] reveal differences owing to the corresponding air cavities and bubble entrainment. The flow focusing is the most efficient for $\Wen = 9$, as distinct from the sharp peak in the transient force evolution. This capillary resonance leads to a strong downward momentum jet and hence the maximum amplitude $F_2$ at time $t = t_2$. Bubble entrainment does not occur for either $\Wen = 5$ [Fig~\ref{Fig_RegimeII}(b-i)] or $\Wen = 12$ [Fig~\ref{Fig_RegimeII}(b-iii)] (see $t_2 < t < 1.2t_2$). Consequently, the maximum force amplitude diminishes for these two cases [Fig.~\ref{Fig_RegimeII}(a-i, a-iii)]. 

\begin{figure}
	\centering
	\includegraphics[width=\linewidth]{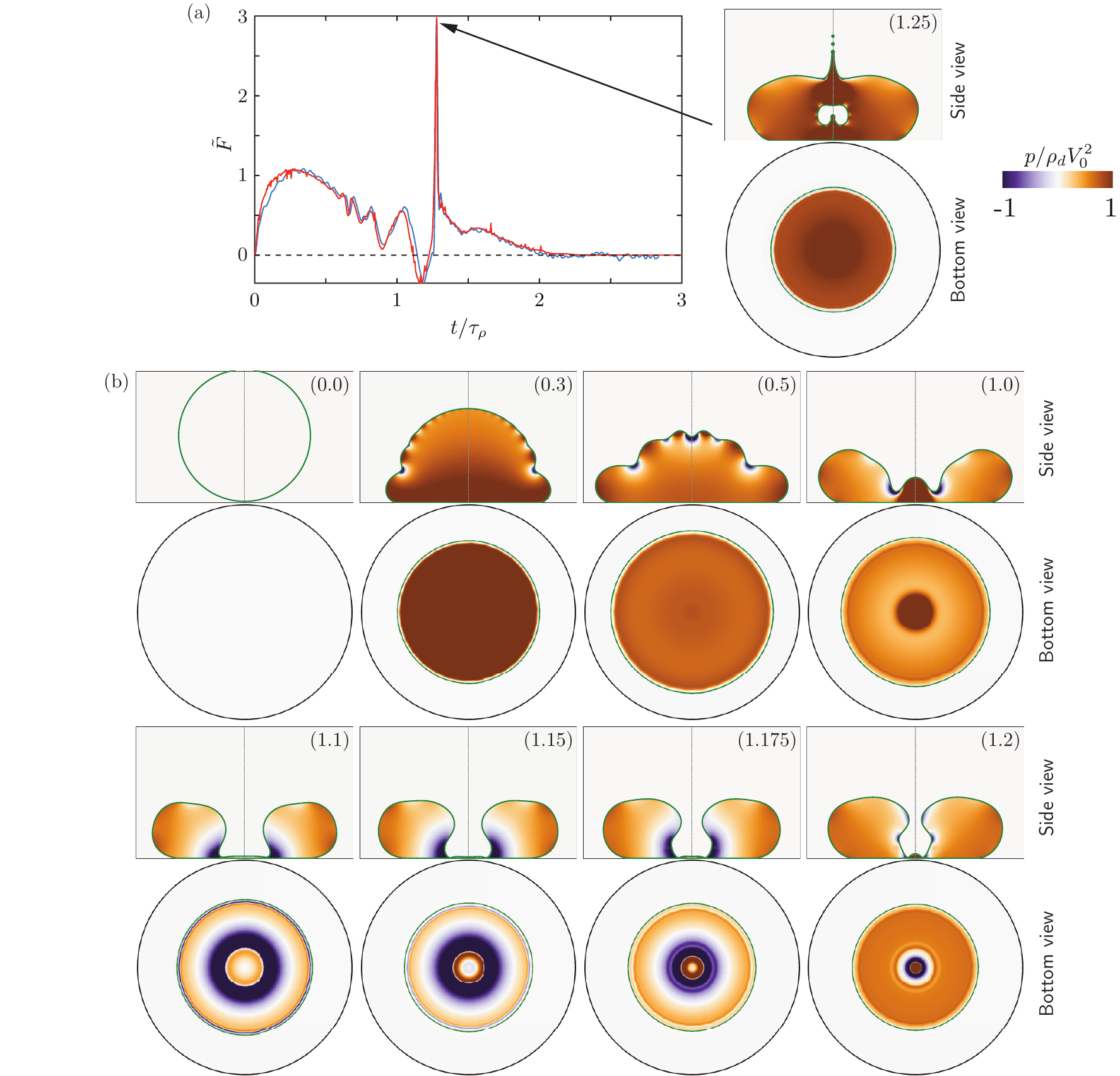}
	\caption{Contact force and pressure filed during drop impact on the superhydrophobic surface in the singular Worthington jet regime, $\Wen = 9$. (a) Temporal variation of the contact force. Notice that the contact force is negative for $1.1\tau_\rho \lessapprox t \lessapprox 1.2\tau_\rho$. (b) Simulation snapshots showing the pressure field $p$ normalized by the inertial pressure $\rho_dV_0^2$ in the side and bottom view images. The numbers in the top right of each numerical snapshot mentions the dimensionless time $t/\tau_\rho$.}
	\label{Fig_Pressure}
\end{figure}

Another characteristic feature of this regime is the occurrence of negative contact force between the drop and the substrate immediately before the formation of a singular Worthington jet and the second peak in normal contact force. Fig.~\ref{Fig_Pressure} illustrates one such case for $\Wen = 9$ where the contact force is negative for $1.1\tau_\rho \lessapprox t \lessapprox 1.2\tau_\rho$ implying that the drop is pulling on the substrate instead of pushing it (Fig.~\ref{Fig_Pressure}(a)). Earlier works \cite{grinspan2010impact, Li2014, Gordillo2018} have attributed this negative force to the wetting properties of the substrates, particularly adhesion between the drop and the substrate \cite{samuel2011study, liimatainen2017mapping}, viscoelastic effects or deformation of the substrate \cite{Gordillo2018}. However, none of these effects are present in our work. To demystify the occurrence of this negative force, we monitor the pressure field inside the drop (side view, Fig.~\ref{Fig_Pressure}(b)) and on the substrate (bottom view, Fig.~\ref{Fig_Pressure}(b)). We observe large negative pressures (purple regions in the pressure field) on the substrate immediately prior to the formation of a singular Worthington jet and the second peak in normal contact force owing to the negative curvature on the surface of the drop as the air-cavity forms due to focusing of the capillary waves. Consequently, negative capillary pressure causes a pressure deficit inside the drop, and the drop pulls on the substrate instead of pushing it (brown regions in the pressure field).

\subsection{Transitional regime II-III and inertial regime III}

\begin{figure}
	\includegraphics[width=\linewidth]{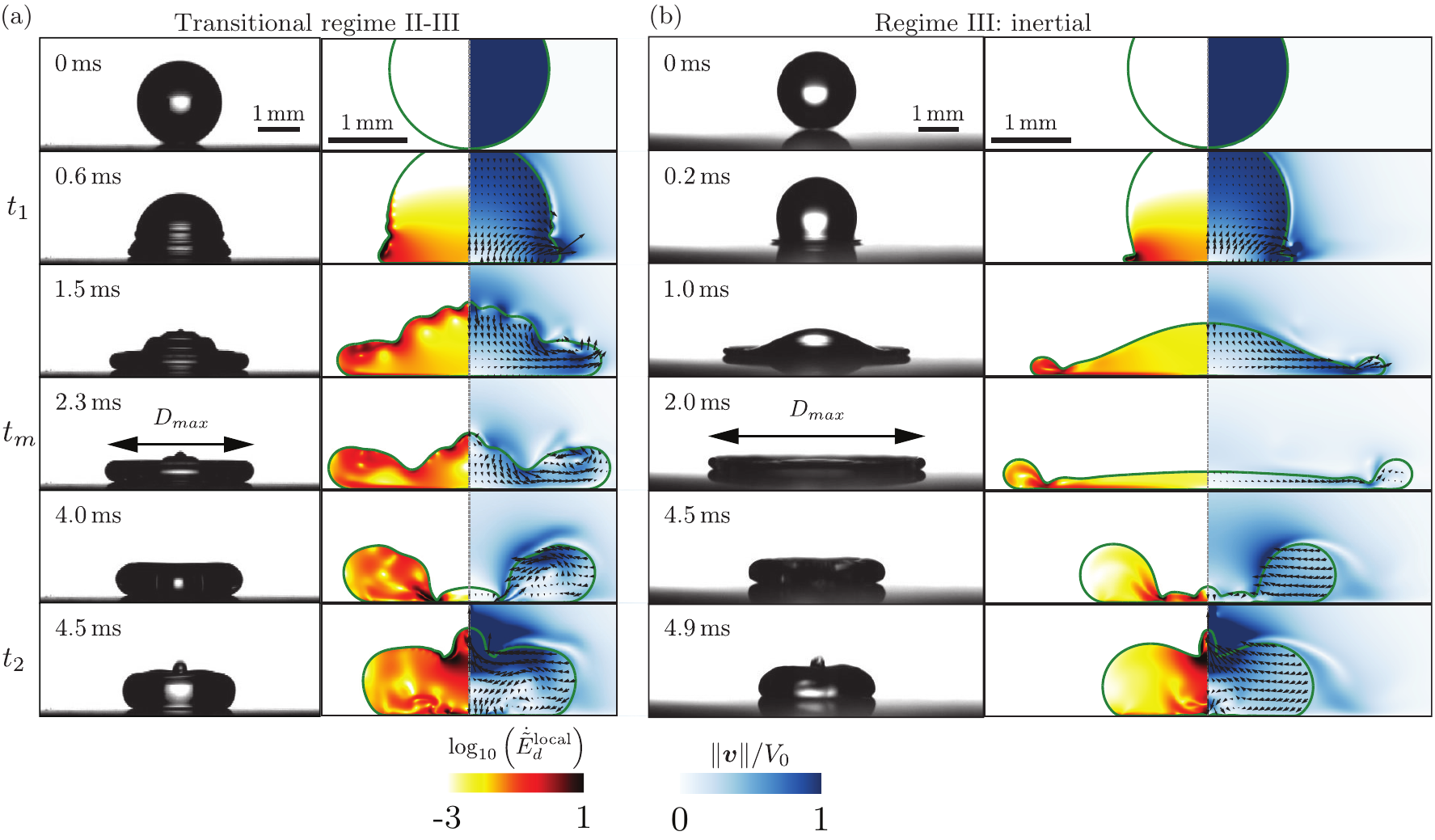}
	\caption{Snapshots of drop shape with time at different Weber numbers, $\Wen = $  (a) 20 (b) 80.}
	\label{Fig_Transition}
\end{figure}

In the main text, we used the flow focusing due to drop retraction to find an expression for the amplitude $F_2$ (also see, Eq.~(\red{2}) of main text), 
\begin{equation}
	\tilde{F}_2 = \frac{F_2}{\rho_d V_0^2D_0^2} \sim \frac{\tilde{v}_j\tilde{v}_2}{\tilde D_2},\tag{2}
	\label{Eqn::TheoryForce2}
\end{equation}
\noindent that entails two scaling behaviors depending on the $\Wen$ (see Eq.~(\red{1}) of main text),
\begin{equation}
	\tilde F_2 = \frac{F_2}{\rho_d V_0^2 D_0^2} \approx
	\begin{cases}
		11\Wen^{-1} &(12.6 < \Wen < 30),\\
		0.37 &(30 < \Wen < 100).
	\end{cases}
	\tag{1}
	\label{eq_1}
\end{equation}
To address the crossover of these two scaling relations, (i.e., $\Wen = 30$, Eq.~(\ref{eq_1})), we check the deformation of the drop at the moment of maximum spreading and the corresponding position of the drop apex. 

To make a comparison, we exemplarily choose $\Wen = 20$ ($D_0 = 2.05\,\si{\milli\meter}, V_0 = 0.83\,\si{\meter}/\si{\second}$) and $\Wen = 80$ ($D_0 = 2.05\,\si{\milli\meter}, V_0 = 1.69\,\si{\meter}/\si{\second}$), and show their impact behaviors in Fig.~\ref{Fig_Transition}(a) and (b), respectively. By simulations, the anatomy of the inner flow field of the drop are discernible (see the right panels). For the case $\Wen = 20$, the solid-liquid contact region is close to the initial drop diameter when $F_1$ is attained at $t_1 = 0.6\,\si{\milli\second}$. Meanwhile, the excited capillary wave propagates along the drop surface and then deforms the drop into a pyramidal shape at $1.5\,\si{\milli\second}$. Then, the drop reaches its maximum spreading diameter $D_{\mathrm{max}}$ at $t_m = 2.3\,\si{\milli\second}$. Notice that at this moment, the drop apex is higher than the height of the rim and is still moving downwards. After that, the drop starts to recoil, and the drop apex descends to its lowest level after $t_m$. During the recoil, the retreating drop deforms into a pancake shape with air in the center, as shown at $4.0\,\si{\milli\second}$. As time progresses, the retracting flow fills the cavity and creates an upward jet at $t_2 = 4.5\,\si{\milli\second}$, which results in $F_2$.

On the other hand, for the case with $\Wen = 80$, a thin liquid film appears, and the solid-liquid contact area is close to the initial drop diameter when $F_1$ is attained at $t_1 = 0.2\,\si{\milli\second}$. Moreover, there is no obvious capillary wave propagating on the drop surface, as shown at $1.0\,\si{\milli\second}$. Then, the drop apex continuously moves downwards, and its height reaches the height of the rim, and this moment happens before the drop reaches its maximum spreading diameter $D_{max}$ at $t_m = 2.0\,\si{\milli\second}$. Shortly after $t_m$, the drop recoils, while the film thickness in the central region remains the same (see $4.5\,\si{\milli\second}$, \cite{Eggers2010}). Then, the thickening retracting flow converges and collides at the film center to form an upward jet to result in $F_2$, as shown at $4.9\,\si{\milli\second}$.

Based on the above results, we gain the following insight. For small $\Wen$ [Fig.~\ref{Fig_Transition}a, $\Wen = 20$], the drop attains $D_{max}$ (at time $t_m$) before its apex descends to its lowest level (at time $\sim$ $D_0/V_0$), leading to a puddle-shaped drop \cite{Wildeman2016}. This observation indicates that at $t_2$, a competition exists between two flows in the central region of the drop, respectively, coming from the rim and the drop apex. However, for large $\Wen$ [Fig.~\ref{Fig_Transition}(b), $\Wen = 80$], the drop apex attains its lowest level before $t_m$, so the drop has a pizza shape \cite{Eggers2010, Wildeman2016}. Equating the two timescales together, one obtains a crossover Weber number $\Wen^* = 25$, which is close to the value $30$ observed in our work. Alternatively, equating the two scaling relations in Eq.~(\ref{eq_1}) gives a more accurate estimate of the crossover Weber number as $\Wen^* = 29.7$.

\subsection{Regime IV: drop splashing}

\begin{figure}
	\includegraphics[width=180mm]{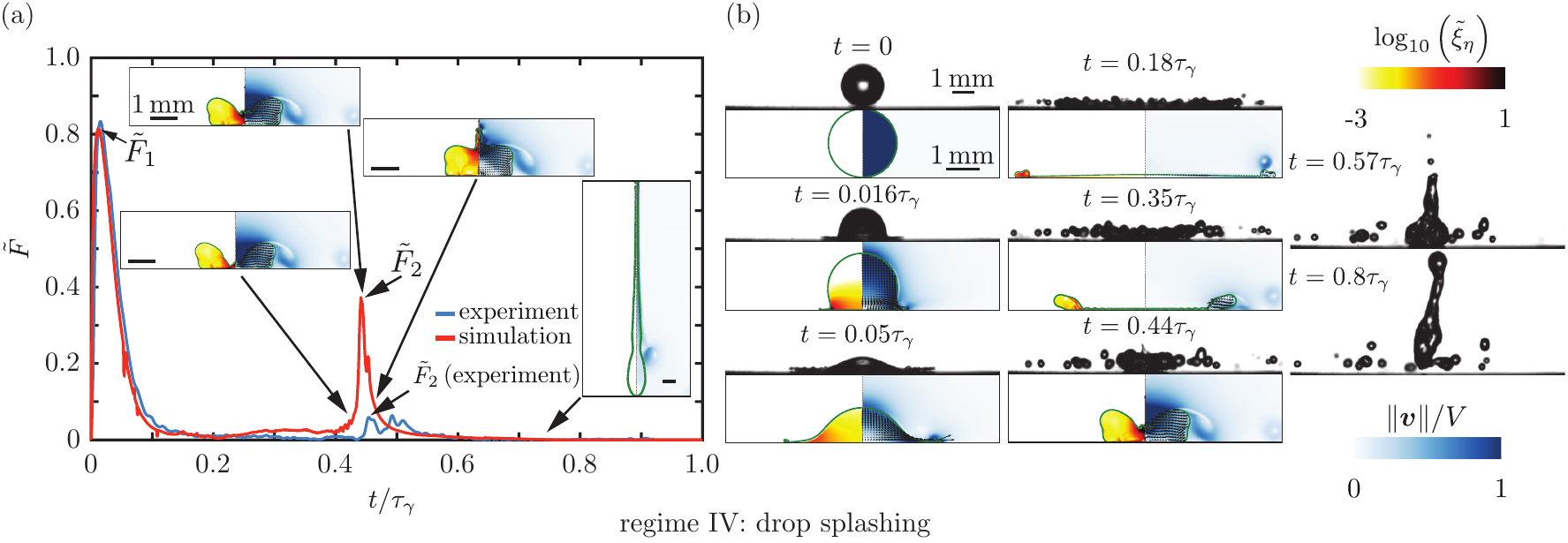}
	\caption{Drop impact on the superhydrophobic surface at a high Weber number, $\Wen = 225$. (a) Evolution of the transient impact force. (b) Snapshots of the corresponding drop geometry in the spreading and recoiling stages. Notice that the experimental and numerical $F(t)$ magnitudes only disagree near $t = t_2$. Further, the time at which the second peak is reached is still at $t_2 \approx 0.44\tau_\gamma$, as explained in the main text. The left part of each numerical snapshot shows the dimensionless local viscous dissipation rates on a $\log_{10}$ scale and the right part shows the velocity field magnitude normalized with the impact velocity. The black velocity vectors are plotted in the center of mass reference frame of the drop to clearly elucidate the internal flow. Also see Supplemental Movie~\red{S6}.}
	\label{Fig_Splashing}
\end{figure}

The main text described the differences between the experimental and numerical observations in regime IV. Here, we further delve into this discrepancy to identify the reasons behind it [Fig~\ref{Fig_Splashing}]. At some high impact velocities (large Weber number), splashing occurs in the experiments \cite{Josserand2016}. At such high $\Wen$, the surrounding gas atmosphere destabilizes the rim, breaking it  \cite{Eggers2010, riboux2014experiments}. Therefore, in regime IV, kinetic and surface energies are lost due to the formation of satellite drops [Fig~\ref{Fig_Splashing}(b)], resulting in diminishing $\tilde{F}_2$ in the experiments [Fig~\ref{Fig_Splashing}(a)]. Obviously, such azimuthal instability is absent in the simulations (axisymmetric by definition), which leads to a better flow-focusing at the center. Consequently, Eq.~(\ref{Eqn::TheoryForce2}) holds only for the simulations in regime IV and not for the experiments. Notice that the experimental and numerical $F(t)$ magnitudes only disagree near $t = t_2$. Further, the time at which the second peak is reached is still at $t_2 \approx 0.44\tau_\gamma$, as explained in the main text. Also see Supplemental Movie~\red{S6}.

\section{Calculation of the jet and retraction characteristics}
\subsection{Experiments}
\begin{figure}
	\includegraphics[width=14cm]{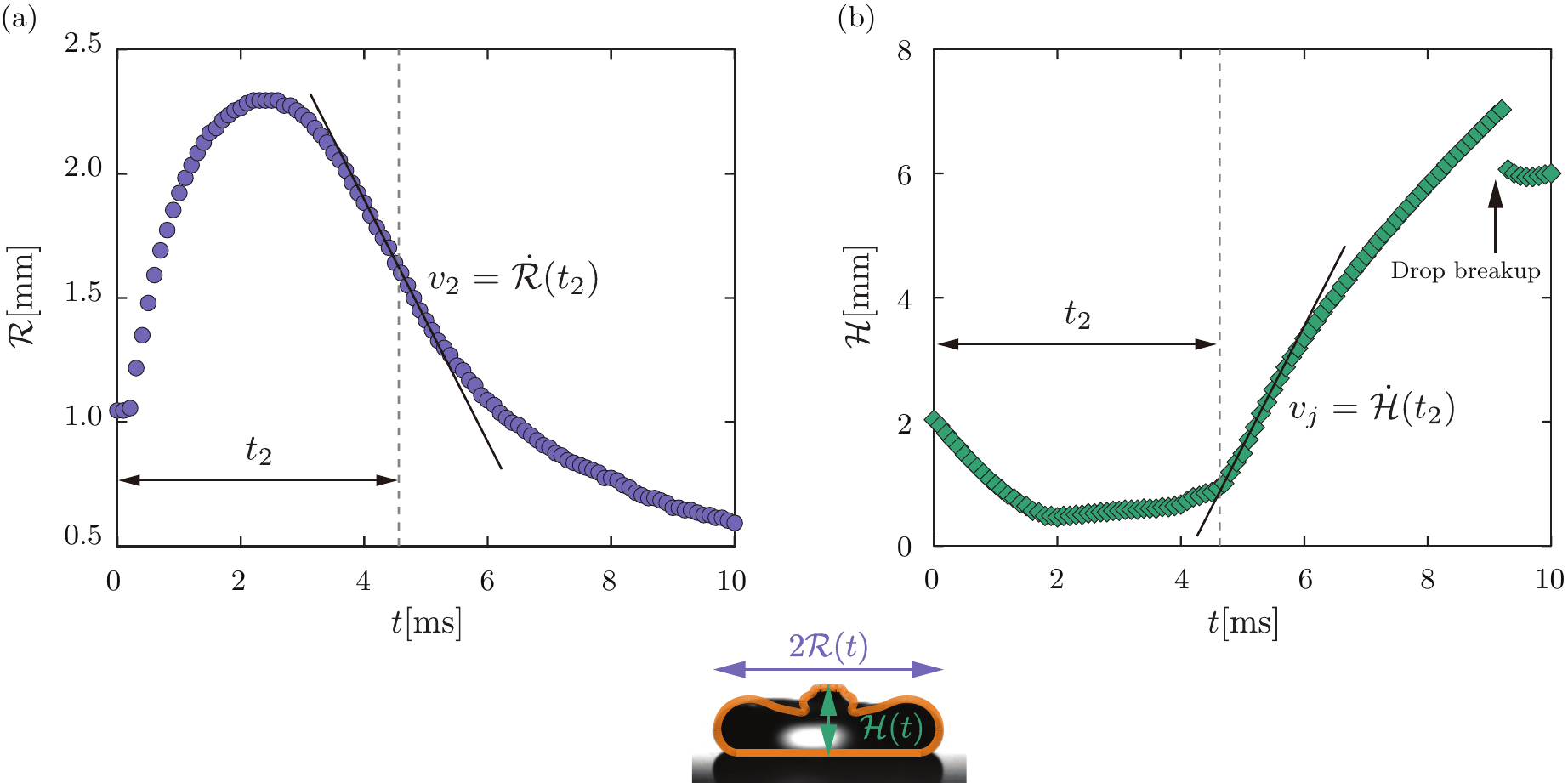}
	\caption{Experimental time evolution of (a) spread radius $\mathcal{R}(t)$ and (b) the drop height $\mathcal{H}(t)$. Inset illustrates the drop geometry. The retraction velocity $v_2 = \dot{\mathcal{R}}(t_2)$ and the jet velocity $v_j = \dot{\mathcal{H}}(t_2)$ are represented by the slopes of the solid lines at $t_2$ in (a) and (b), respectively. Here, $\Wen = 40$.}
	\label{Fig_ExpU2Uj}
\end{figure}

In this section, we illustrate how to extract $v_2$ and $v_j$ from the experiments. As an example, we choose $\Wen = 40$. As shown in Fig.~\ref{Fig_ExpU2Uj}, we first track the instantaneous values of the height $\mathcal{H}(t)$ and the width $2\mathcal{R}(t)$ of the drop. From Fig.~\ref{Fig_ExpU2Uj}, we can see that just after the impingement, the drop height decreases with a constant velocity \cite{Eggers2010, Gordillo2018}. As time progresses, $\mathcal{R}(t)$ and $\mathcal{H}(t)$ respectively reach their maximum and minimum values simultaneously (at $t_{\mathrm{m}} \approx 2.5\,\si{\milli\second}$). Moving forward in time, $\mathcal{R}(t)$ decreases, whereas $\mathcal{H}(t)$ increases linearly until $t_2$. After this moment, we observed a sharp increase of $\mathcal{H}(t)$ until the drop breaks into the base drop and a small droplet (see the inset in Fig.~\red{1}(c) of the main text). We define the recoiling velocity $v_2$ of the drop and the jet velocity $v_j$ as:

\begin{align}
	v_2 = \dot{\mathcal{R}}(t_2) &= \left. \frac{\mathrm{d}\mathcal{R}(t)}{\mathrm{d}t} \right|_{t_2},\\
	v_j = \dot{\mathcal{H}}(t_2) &= \left. \frac{\mathrm{d}\mathcal{H}(t)}{\mathrm{d}t} \right|_{t_2}.
\end{align}

As shown in Fig.~\ref{Fig_ExpU2Uj}(a), $v_2= \dot{\mathcal{R}}(t_2)$ is obtained by a linear fitting (black line) to the experimental data around $(t_2, \mathcal{R}(t_2))$. Similarly, as shown in Fig.~\ref{Fig_ExpU2Uj}(b), $v_j = \dot{\mathcal{H}}(t_2)$ is obtained by a linear fitting (black line) to the experimental data around $(t_2, \mathcal{H}(t_2))$. Note that in the experiments, we can only measure the maximum height of the drop. Consequently, when the rim thickness exceeds the drop's height, $\mathcal{H}$ identifies the height of the rim. So, we use the datapoints after $t = t_2$ to calculate the jet velocity. Nonetheless, the jet velocity $v_j$ extracted at $t = t_2^{+}$ from the experiments are consistent very well with our simulation (where we can precisely calculate the jet velocity, see \S~\ref{sec::calculations Sim}), as well as the results obtained by \citet{bartolo2005retraction}, as discussed in \S~\ref{sec::calculations results}.

\subsection{Simulations}
\label{sec::calculations Sim}
\begin{figure}
	\includegraphics[width=\linewidth]{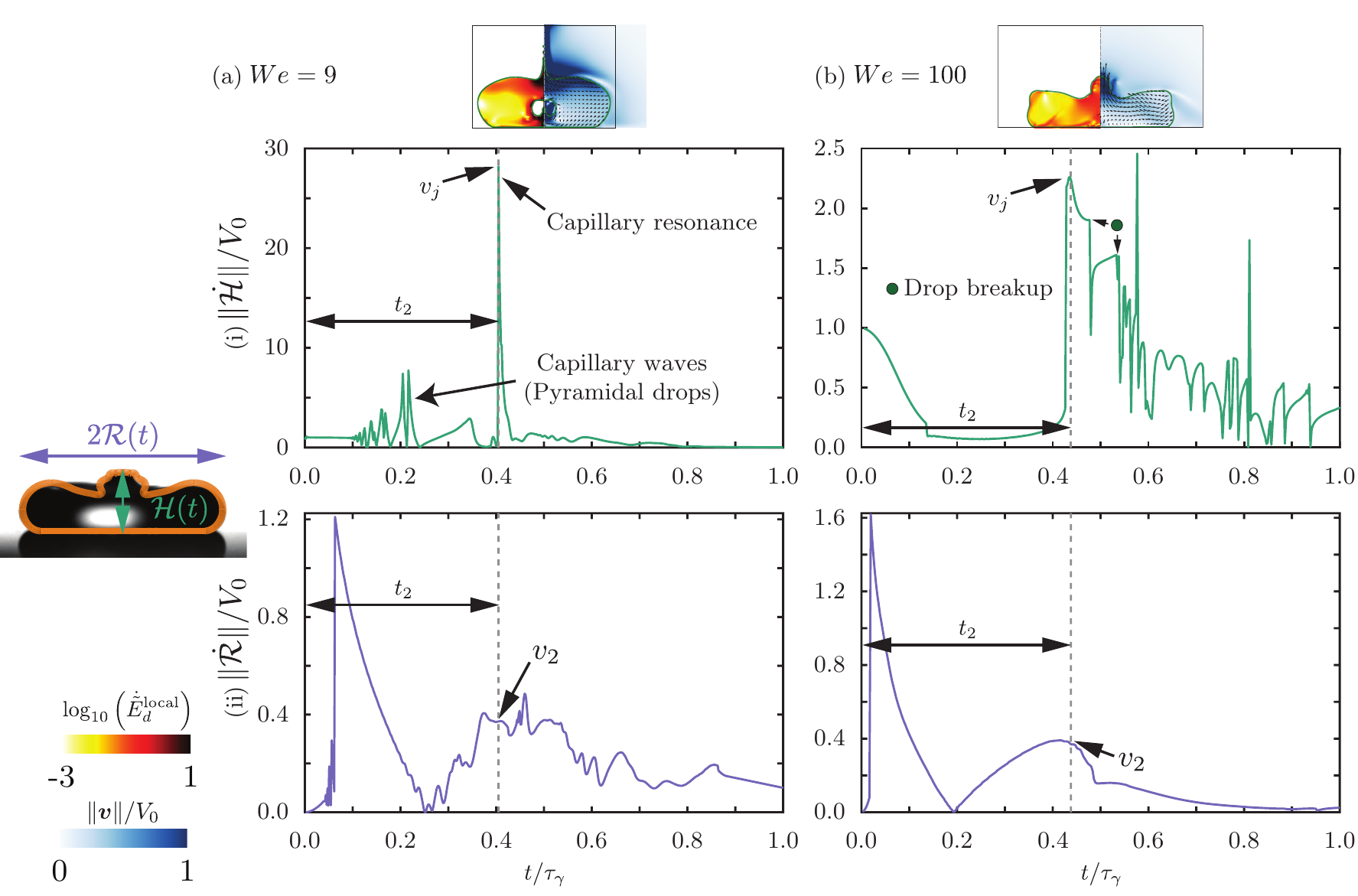}
	\caption{Calculation of the (i) jet velocity ($v_j$) and (ii) retraction velocity ($v_2$) for Eq.~(\red{2}) of the main manuscript for two representative cases: $\Wen =$ (a) $9$ and (b) $100$. Inset illustrates the drop geometry where $\mathcal{H}$ is the height of the drop at the axis of symmetry and $2\mathcal{R}$ is its radial extent. The jet velocity is $v_j = \dot{\mathcal{H}}(t_2)$ and the retraction velocity is $v_2 = \dot{\mathcal{R}}(t_2)$. Notice that the time at which second peak is reached still scales with the inertial-capillary timescale $t_2 \sim \tau_\gamma$, as described in the main text, irrespective of the $\Wen$ ($t_2 = 0.405\tau_\gamma$ for $\Wen = 9$, and $t_2 = 0.437\tau_\gamma$ for $\Wen = 100$).}
	\label{Fig_calculations}
\end{figure}
To characterize the jet, we track the interfacial location (or height of the drop, $\mathcal{H}(t)$) at the axis of symmetry ($r = 0$). Similarly, to characterize retraction, we track the radial extent of the drop ($2\mathcal{R}(t)$). Further, $\dot{\mathcal{H}} = \mathrm{d}\mathcal{H}/\mathrm{d}t$ measures the velocity of this jet, and $\dot{\mathcal{R}} = \mathrm{d}\mathcal{R}/\mathrm{d}t$ accounts for the retraction velocity. Fig.~\ref{Fig_calculations} shows the temporal variation of $\dot{\mathcal{H}}$ and $\dot{\mathcal{R}}$ for two representative Weber numbers, $\Wen = 9\,\text{and}\,100$. As the drop impacts, the top of the drop keeps moving with a constant velocity ($\dot{\mathcal{H}} \approx V_0$) \cite{Eggers2010, Gordillo2018}. However, during this period, the radial velocity magnitude increases to a maximum and then decreases to zero at the instant of maximum spreading. 

For low to moderate Weber number impacts ($\Wen = 9$ in Fig.~\ref{Fig_calculations}a), the pyramidal morphology result in capillary oscillations aiding the flow focusing in the retraction phase. Consequently, both the normal force ($F$) and the jet velocity ($\dot{\mathcal{H}}$) reach the maxima simultaneously at $t = t_2$. Further, the retraction velocity show oscillations due to capillary waves.

On the other hand, for high Weber number impacts ($\Wen = 100$ in Fig.~\ref{Fig_calculations}b), the jet velocity is minimum at the instant of maximum spreading. Then, Taylor-Culick type retraction occurs increasing the retraction velocity to a maximum which then decreases due to finite size of the drop \cite{bartolo2005retraction, Eggers2010, pierson2020revisiting, deka2020revisiting}.  During this retraction phase, flow focusing and asymmetry provided by the substrate lead to a sudden increase in the jet velocity that is immediately followed by occurrence of the second peak in the transient force profile (at $t_2$). The retraction velocity at this instant is very close to its maximum temporal value. 

For both cases, notice that the time at which second peak is reached still scales with the inertial-capillary timescale $t_2 \sim \tau_\gamma$, as described in the main text, irrespective of the $\Wen$ ($t_2 = 0.405\tau_\gamma$ for $\Wen = 9$, and $t_2 = 0.437\tau_\gamma$ for $\Wen = 100$).

In summary, 
\begin{align}
	v_j &= \dot{\mathcal{H}}(t_2)\\
	v_2 &=  \dot{\mathcal{R}}(t_2).
\end{align}

Lastly, we can also characterize the maximum lateral extent $D_2$ of the drop at the instant $t_2$ of second peak in the normal reaction force $F_2$ as

\begin{align}
	D_2 = 2\mathcal{R}\left(t_2\right),
\end{align}

\noindent in both experiments as well as simulations. 

\subsection{Results}
\label{sec::calculations results}
\begin{figure}
	\includegraphics[width=\linewidth]{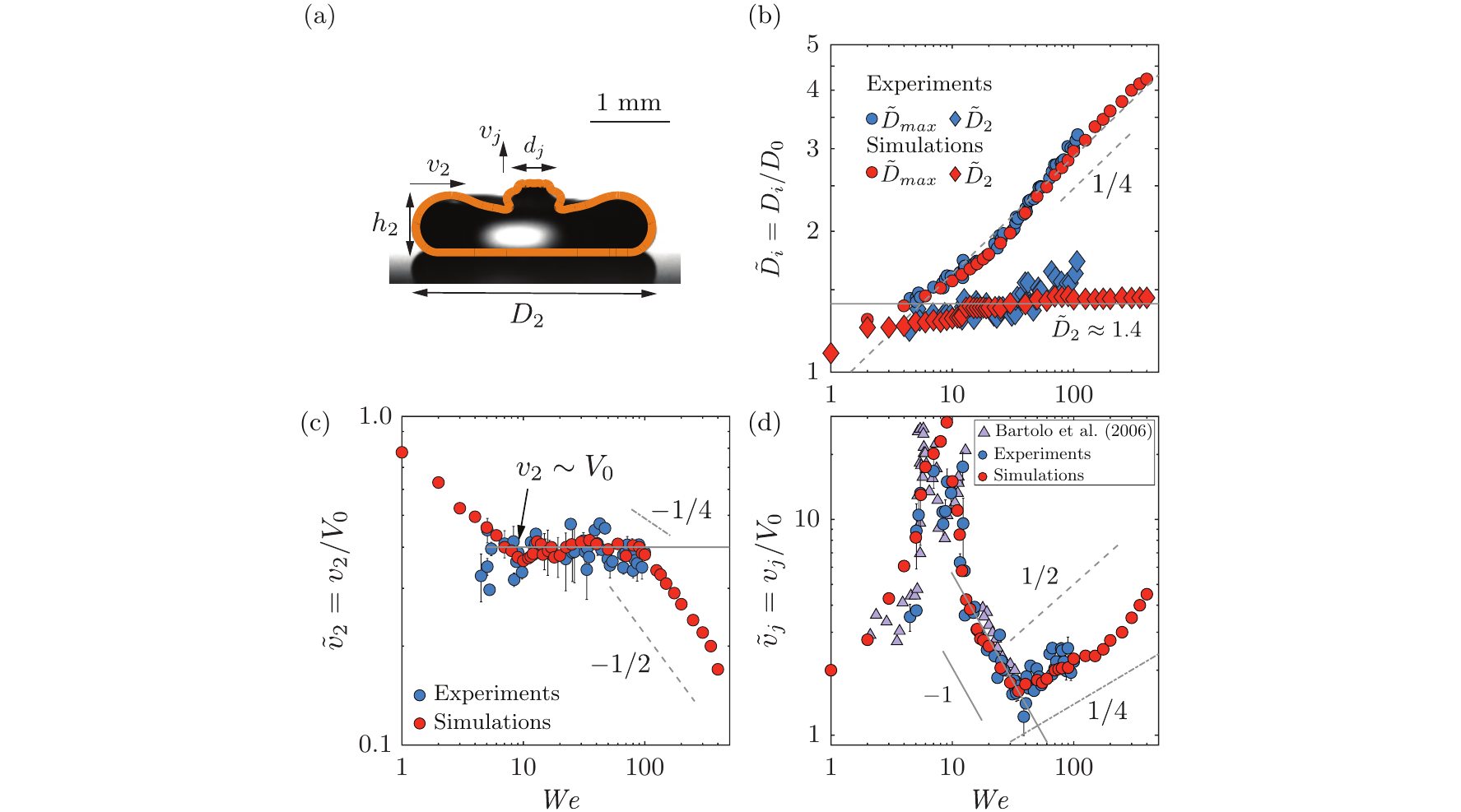}
	\caption{(a) Drop geometry at $t_2$ for $\Wen = 40$ (along with the drop contour from numerics in orange) to illustrate the drop spreading diameter $D_2$, drop height $h_2$, retraction velocity $v_2$, jet diameter $d_j$ and jet velocity $v_j$. (b) Variation of the dimensionless spreading diameter at $t_m$ and $t_2$ (given by $\tilde D_{max} = \tilde D (t_m)  $ and $\tilde D_2 = \tilde D(t_2)$, respectively). The gray dotted and solid lines represent $\tilde D_{max} \sim \Wen^{1/4}$ and $\tilde D_2 \approx 1.4$, respectively. (c) Variation of drop retraction velocity $\tilde{v}_2$ at $t_2$ with $\Wen$. The gray solid line represents $v_2 \sim V_0$. The gray dotted and dashed-dotted lines correspond to $\Wen^{-1/2}$ and $\Wen^{-1/4}$, respectively,  and are meant only as guides to the eye. (d) Variation of the (non-dimensionalized) jet velocity $\tilde{v}_j = v_j/V_0$ with $\Wen$. The data from \citet{Bartolo2006Singular} are also shown in the same panel. The gray solid line represents $\tilde{v}_j \sim \Wen^{-1}$. The gray dotted and dashed-dotted lines correspond to $\Wen^{1/2}$ and $\Wen^{1/4}$, respectively, and are meant only as guides for the eye.}
	\label{Fig_details}
\end{figure}
We will devote the rest of this supplemental material to relate the different flow properties in Eq.~(\ref{Eqn::TheoryForce2}) to the control parameter, i.e., the impact Weber number $\Wen$. For the transitional regime II-III ($12.6 < \Wen < 30$), at the moment of second peak, the dimensionless diameter ($\tilde{D}_2$) and the dimensionless drop retraction velocity ($\tilde{v}_2$) are independent of the impact Weber number $\Wen$ [Figs.~\ref{Fig_details}(b, c)]. Further, the jet velocity decreases with increasing Weber number following $\tilde{v}_j = v_j/V_0 \sim 1/\Wen$ [Figs.~\ref{Fig_details}(d)]. This decrease is consistent with the data extracted from \citet{Bartolo2006Singular}. Substituting these in Eq.~(\ref{Eqn::TheoryForce2}), one obtains $\tilde{F}_2 \sim 1/\Wen$. However, the prefactor that best fits the experimental and numerical data in Eq.~(\ref{eq_1}) is much larger than order 1, which may be caused by the enhanced flow and momentum focusing due to both capillary waves and drop retraction.

For regime III ($30 < \Wen < 100$), there is a slight increase in $\tilde{D}_2$ [Fig.~\ref{Fig_details}(b)] but it is still best represented by a plateau. Furthermore, with increasing $\Wen$, $\tilde{v}_2$ decreases whereas $\tilde{v}_j$ increases [Figs.~\ref{Fig_details}(c,d)]. We have shown gray lines as guides to the eye to represent these trends. However, due to limited range of $\Wen$, we refrain from claiming any scaling relations here. Coincidentally, these changes in $\tilde{D}_2$, $\tilde{v}_2$, and $\tilde{v}_j$ compensate each other such that Eq.~(\ref{Eqn::TheoryForce2}) still holds. Consequently, the second peak force scales with the inertial pressure force $F_2 \sim \rho_d V_0^2D_0^2$ [Eq.~(\ref{eq_1})].

Alternatively, we can use the expressions for the amplitude of the second peak of force between the drop and the substrate to predict the velocity of the Worthington jet. For $\Wen \gg 1$, the drop forms a thin sheet at the instant of maximum spreading ($t = t_m$, see Fig.~\ref{Fig_Transition}(b) and~\ref{Fig_Splashing}). This sheet retracts following Taylor-Culick type retraction at low Ohnesorge numbers ($\Ohn \ll 1$). As a result, the retraction velocity scale can be given as \cite{bartolo2005retraction, Eggers2010}:

\begin{align}
	v_2 \sim \sqrt{\frac{\gamma}{\rho_d h_2}},
	\label{Eqn::TC}
\end{align}

\noindent where, $\gamma$ and $\rho_d$ are the surface tension coefficient and density of the liquid drop, respectively. Further, $v_2$ is the retraction velocity at $t = t_2$ and $h_2$  is the height of the drop at that instant which is related to the spreading diameter following volume conservation as $h_2 \sim D_0^3/D_2^2$. Substituting this expression in Eq.~(\ref{Eqn::TC}) and normalizing with $V_0$, we get

\begin{align}
	\tilde{v}_2 \sim D_2\sqrt{\frac{\gamma}{\rho_dV_0^2 D_0^3}} = \frac{\tilde{D}_2}{\sqrt{\Wen}}.
	\label{Eqn::TC2}
\end{align}

\noindent The finite size of the retracting drop may account for the deviations from Eq.~(\ref{Eqn::TC2}) in Fig.~\ref{Fig_details}(b,c) \cite{pierson2020revisiting, deka2020revisiting}. Further, using $\tilde{F}_2 \sim \tilde{v}_2\tilde{v}_j/\tilde{D}_2 \sim \mathcal{O}\left(1\right)$ for $\Wen \gg 1$, we obtain

\begin{align}
	\tilde{v}_j \sim \sqrt{\Wen}.
	\label{Eqn::TCj}
\end{align}

\noindent Unfortunately, we cannot confirm the validity of this scaling behavior in Fig.~\ref{Fig_details}(d) due to a limited range of $\Wen$ as mentioned above. For completeness, in Fig.~\ref{Fig_details}(b) we also show the maximum spreading diameter from our experiments and simulations are in a remarkable agreement. We refer the readers to \cite{Clanet2004, Eggers2010, laan2014maximum} for further discussions on the influence of $\Wen$ on the maximum spreading diameter.


\noindent {\it Outlook on the scaling relations:} In this section and Fig.~\ref{Fig_details}, we probe several scaling behaviors in an attempt to relate the internal flow characteristics, i.e., the jet and retraction velocities, to the impact Weber number. However, verifying the predicted scaling behaviors requires a larger range of Weber numbers that we do not study due to experimental and numerical limitations. For example, at very high-velocity impacts, the drop splashes and breaks into many satellite droplets in the experiments \cite{riboux2014experiments}. For future work, we suggest that one could experimentally probe this regime by suppressing the azimuthal instability (for instance, by reducing the atmospheric pressure \cite{xu2005drop}). However, even in such a scenario, at very high impact velocities ($\Wen \gg 1$), the substrate roughness may play a role in both drop spreading and retraction \cite{visser2015dynamics}. In numerical simulations, one can probe this regime by using drops that are slightly more viscous than water, as done by \citet{Eggers2010}. However, such a study is numerically costly for water drops impacting at very large $\Wen$ due to the separation of length scales between the initial diameter of the drop and the very thin lamella during spreading. Furthermore, the interfacial undulations (traveling capillary waves) further restrict both the spatial and temporal resolutions.

\section{Supplemental movie captions}

\begin{enumerate}
	\item \textbf{Supplemental movie 1.} The evolution of the transient force of a water drop impacting on the superhydrophobic surface at a moderate Weber number $\Wen = 40$ (corresponding to $D_0 = 2.05\,\si{\milli\meter}$ and $V_0 = 1.20\,\si{\meter}/\si{\second}$), with simultaneous drop geometry recorded experimentally at 10,000 fps with the exposure time of 1/20,000 s. The left part of the numerical video shows the dimensionless local viscous dissipation rates on a $\log_{10}$ scale and the right part shows the velocity field magnitude normalized with the impact velocity.
	\item \textbf{Supplemental movie 2.} The evolution of the transient force of a water drop impacting on the superhydrophobic surface at a low Weber number $\Wen = 2$ (Regime I: capillary), with simultaneous drop geometry evolution. The left part of the numerical video shows the dimensionless local viscous dissipation rates on a $\log_{10}$ scale and the right part shows the velocity field magnitude normalized with the impact velocity.
	\item \textbf{Supplemental movie 3.} The evolution of the transient force of a water drop impacting on the superhydrophobic surface at Weber number $\Wen = 9$ (corresponding to $D_0 = 2.76\,\si{\milli\meter}$ and $V_0 = 0.49\,\si{\meter}/\si{\second}$), with simultaneous drop geometry recorded experimentally at 10,000 fps with the exposure time of 1/50,000 s. The ultra-thin and fast singular jet is reminiscent of the hydrodynamic singularity. The left part of the numerical video shows the dimensionless local viscous dissipation rates on a $\log_{10}$ scale and the right part shows the velocity field magnitude normalized with the impact velocity.
	\item \textbf{Supplemental movie 4:} Boundaries of the singular jet regime: the evolution of the transient force of a water drop impacting on the superhydrophobic surface at three representative Weber numbers in regime II, $\Wen = 5$, $\Wen = 9$, and $\Wen = 12$, with simultaneous drop geometry evolution. The left part of the numerical video shows the dimensionless local viscous dissipation rates on a $\log_{10}$ scale and the right part shows the velocity field magnitude normalized with the impact velocity.
	\item \textbf{Supplemental movie 5:} Contact force and pressure filed during drop impact on the superhydrophobic surface in the singular Worthington jet regime, $\Wen = 9$. Simulation shows the pressure field $p$ normalized by the inertial pressure $\rho_dV_0^2$ on the left and the magnitude of velocity field $\|\boldsymbol{v}\|$ normalized by the impact velocity $V_0$ on the right in the side view images and pressure field in the bottom view images.s
	\item \textbf{Supplemental movie 6:} The evolution of the transient force of a water drop impacting on the superhydrophobic surface at a high Weber number $\Wen = 225$ (drop splashing regime, corresponding to $D_0 = 2.05\,\si{\milli\meter}$ and $V_0 = 2.83\,\si{\meter}/\si{\second}$), with simultaneous drop geometry recorded experimentally at 10,000 fps with the exposure time of 1/20,000 s. The left part of the numerical video shows the dimensionless local viscous dissipation rates on a $\log_{10}$ scale and the right part shows the velocity field magnitude normalized with the impact velocity.
\end{enumerate}

\bibliography{supplement}